\documentclass[preprint,english,aps,pre]{revtex4-2}
\usepackage[utf8]{inputenc}
\usepackage{amsmath}
\usepackage{graphicx}
\usepackage{booktabs}
\usepackage{orcidlink}
\usepackage{subcaption}
\usepackage[labelsep=space]{caption}
\usepackage{xcolor}
\usepackage{amssymb}
\usepackage{babel}
\usepackage{xcolor}

\usepackage[normalem]{ulem}

\makeatother

\begin{document}

\title{Exploring new speciation patterns: architecture of genetic similarity networks}

\author{David M. Schneider$^{\dagger,**}$,\orcidlink{0000-0002-0124-8759}}
\author{Claudio Pastorino$^{*,\ddagger}$,\orcidlink{0000-0002-4833-5999}}

\affiliation{$^{\dagger}$Grupo de Física Estadística e Interdisciplinaria, Centro
Atómico Bariloche, CNEA, Av.~Bustillo 9500, 8400  Pcia.~de Río Negro,  Argentina}

\affiliation{$^{**}$Consejo Nacional de Investigaciones Científicas y Técnicas (CONICET) Av.~de los Pioneros 2350, 8400  Pcia.~de  Río Negro, Argentina}

\affiliation{$^{*}$Departamento de Física de la Materia Condensada, Centro
Atómico Constituyentes, CNEA, Av. Gral.~Paz 1499, 1650 Pcia.~de Buenos
Aires, Argentina}

\affiliation{$^{\ddagger}$Instituto de Nanociencia y Nanotecnología, CONICET-CNEA, CAC.}

\begin{abstract}

We use an individual based model to identify signatures of speciation modes in genetic similarity networks. These networks group individuals based on their genetic similarity, defined through Hamming distances between binary strings representing their genomes.

The results of our extensive simulations demonstrate that sympatric and parapatric speciation modes are clearly distinguishable and measurable  through specific network metrics. In particular, structural properties —including clustering coefficient, mean degree, geodesic distance, and assortativity— show significant differences in both stationary values and transient phases, giving rise to a novel framework for distinguishing between speciation modes.

\end{abstract}
\maketitle
\section{Introduction}

Speciation, i.e., the generation of new biological species from an ancestral population, is a complex process that can occur through different paths. Essentially, it involves reproductive and ecological barriers that separate an initially homogeneous set of individuals into reproductively isolated groups. The elucidation of the mechanisms which promote speciation, constitutes a central problem in evolutionary biology, as their study is essential for understanding the genesis of biodiversity \cite{Coyne}. According to the characteristics of the segregation process, one recognizes allopatric speciation, in which physical barriers lead to the formation of non-overlapping species, sympatric speciation, in which the new formed species share a common area of distribution, and a spectrum of intermediate scenarios of partial superposition, collectively referred to as ‘parapatric’.
 Allopatric speciation is the predominant mode in nature, supported by comparative analyses in Drosophila \cite{Coyne1989patterns,Coyne1997patterns} and in plants such as palms and conifers \cite{olivares2025likelihood}. Sympatric speciation, although controversial, has well-documented examples (such as crater lake cichlids \cite{Schliewen1994,Kautt2022} or host races of Rhagoletis \cite{Bush1969,Feder1988,Smith1988}), but is considered rare and requires highly restrictive ecological and genetic conditions: strong disruptive selection, assortative mating, and, in many models, pleiotropy (the influence of a gene in the  other seemingly unrelated phenotypic traits) between ecological and mating traits \cite{bolnick2007sympatric, kondrashov_mina_1986, kondrashov_shpak_1998,gavrilets2006maynard}.
 In this context, theoretical work by Udovic  established that frequency-dependent selection and disruptive selection at a polymorphic equilibrium can lead to the evolution of reproductive isolation through epistatic interactions between loci \cite{udovic1980frequency}. Kirkpatrick and Ravigné \cite{kirkpatrick2002speciation} synthesized models of speciation, identifying five key elements (disruptive selection, isolation mechanism, selection type, genetic basis, and initial conditions) which show that geographical context can be viewed as a form of assortative mating, and ultimately establishing  a link between  geographical and ecological influences on sympatric and parapatric regimes. However, Kirkpatrick and Nuismer \cite{kirkpatrick2004sexual} demonstrated that assortative mating may often generate stabilizing sexual selection, which can substantially impede sympatric speciation.
 Age-range correlation (ARC) analyses yield ambiguous results due to post-speciation biogeographic changes, prompting recent approaches to assess the frequency of conditions that facilitate sympatry (such as ``magic traits'' or weak costs of choosiness) rather than counting past events \cite{bolnick2007sympatric}.  Overall, the evidence supports a pluralistic view: allopatry is dominant, but sympatry and parapatry are  possible and likely more common in certain groups or under specific ecological conditions.

The nature of the process or processes giving rise to the present species  are supposed to have profound implications for predicting macroevolutionary patterns, interpreting the phylogenetic and ecological structure of communities, and anticipating species responses to environmental changes such as habitat fragmentation \cite{Rosenzweig,Gavrilets}. As Coyne and Orr emphasize, ``speciation is the process that fills the world with the biological diversity we see'' \cite{Coyne}. Therefore,  unraveling its mechanisms is fundamental to reconstructing the history of life. However, as empirical studies analyzing the geographic overlap between sister species demonstrate \cite{Coyne,Fitzpatrick}, inferring the mode of speciation from contemporary patterns is overwhelmingly difficult, as historical signals can be masked by secondary dispersal or extinctions \cite{Coyne,losos2003phylogenetic}. Overcoming this difficulty requires integrating evidence from phylogeography, comparative genomics, ecology, and the fossil record, converging in what has been termed the "integrated" or "pluralistic" study of speciation \cite{nosil2012ecological,schluter2009evidence}.  Ultimately, understanding speciation modes is a key piece in unveiling the rules that govern the diversification of life on Earth.

In previous simulation studies of an individual-based speciation model, Costa {\it et al.} established a connection between sympatric and parapatric speciation modes and the topology of phylogenetic trees associated with the corresponding evolutionary processes \cite{costa_1, costa_2}. These two speciation modes differ in the extent of gene flow: completely unrestricted in sympatric speciation versus moderately restricted in parapatric speciation. However, the most striking aspect of the parapatric speciation, which distinguishes it from the classical allopatric speciation, is the absence of physical barriers despite limited gene flow.
By constructing  phylogenetic trees from the history of the processes of speciation, the authors computed topological properties such as tree balance and diversification rates, demonstrating that sympatric and parapatric speciation can be identified from tree morphology by analyzing these properties.
Specifically, balanced trees (symmetrical) arise from sympatric speciation with high gene flow and large genomes, whereas asymmetric trees (unbalanced) result from parapatric speciation with low gene flow and small genomes, reflecting uniform or biased diversification patterns, respectively.
All studies were performed by implementing the de Aguiar model \cite{modelo_de_aguiar}, in which individuals are represented by  binary ``genomes'' and populations evolve according to  a reproductive  rule, based on genetic and spacial proximity criteria.
In this work, we undertake an analogous analysis, but focusing on the so-called  `genetic similarity networks' (GSN)\cite{entropy_2025}. These networks are constructed by assigning a node to each individual in the population and a link to each pair of individuals with  compatible  genomes, as detailed below.

The article is organized as follows. In Section \ref{sec:models} we present our version of the de Aguiar model, indicating the specific regions of the parameter space under study. We also compare the regions to those studied by Costa {\it et al.} \cite{costa_1,costa_2} and describe  the construction of the GSN. In  Section \ref{sec:simulations} we report simulation results and structural properties of the networks, both in stationary and transient phases. Conclusions and final discussion are reserved for Section \ref{sec:conclusions}.

\section{Speciation model and network construction}\label{sec:models}

We study an adapted version of the de  Aguiar model in which individuals remain immobile in their spatial positions, unlike the original framwork, in which descendants might diffuse from their parents locations. The details of the de  Aguiar model can be found elsewhere \cite{costa_1, costa_2,modelo_de_aguiar, diploid_vs_haploid, sex_separation}. Here we will explain its essential aspects and the characteristics of our model modifications.

\begin{figure}
\includegraphics[width=0.95\columnwidth]{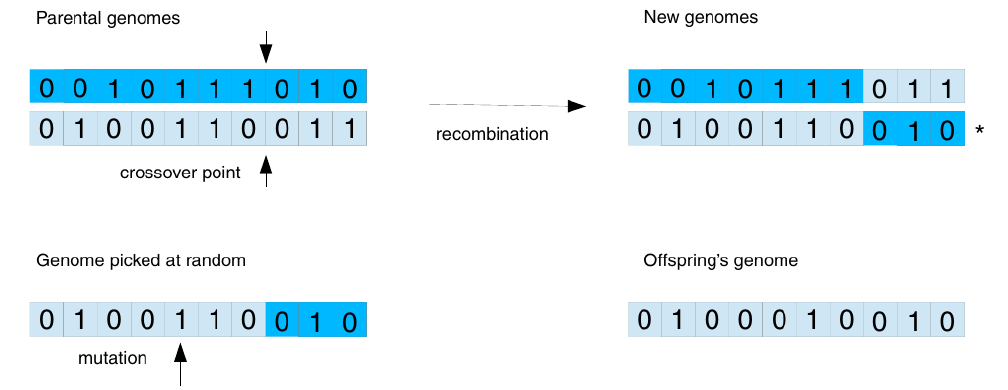}
\caption{Diagram outlining the events involved in reproduction. Top-left: sequences before recombination and  selection of the cross-over point. Top-right: sequences after recombination and selection of a daughter sequence (offspring).
 Bottom-left: offspring before mutation. Bottom-right: offspring after mutation.}
\label{figura_empalme}
\end{figure}

The de Aguiar model comprises a set of $N$ binary sequences of length $B$ (representing the genomes of ``hermaphrodic'' individuals), which are updated according to the following rule: a binary sequence (``mother'')  attempts to mate with a second sequence (“father”),  randomly  selected within a neighborhood defined by a cut-off radius $S$.  If the Hamming distance between the two sequences, defined as the number of unmatched bits, is less than a predetermined threshold defined by a ‘similarity’ parameter $G$, mating occurs. This process consists on the random selection of a recombination point and a subsequent splicing (see Figure \ref{figura_empalme}). One of the two daughter sequences  is randomly selected to replace the mother sequence, and after this event, each bit of  the selected daughter sequence is subject to mutation with probability $\mu$. If the Hamming distance between  mother and father sequences exceeds the threshold $G$, the latter is discarded and a new father is randomly selected within the neighborhood \footnote{The terms `mother' and `father' are used here to distinguish the individual to be replaced in the next generation, from its couple. As in the original model by de  Aguiar, all individuals can mate freely with one another.  A version of the de Aguiar model with true sex separation was studied in \cite{sex_separation}.}. If none of the potential fathers  is compatible according to the  criterion based on the Hamming distance, the mother sequence is discarded and replaced by a new mother  selected at random. This fact implies that some mothers reproduce more than once, creating possible regions of high density of individuals. This procedure is repeated for all the available mothers, giving rise to the offspring generation, which in turn will reproduce in the next time step of the simulation.
In our simulations, we use a variable number of time steps (or ``generations'') and take samples
every certain number of generations. The spatial domain consists of a square with sides of 100 units, in which individuals are distributed in random positions. We use periodic boundary conditions for the search of the mate, within the cut-off radius $S$.  This spatial configuration, with individuals placed at random positions and connected through a cutoff radius, defines the so called  Random Geometric Graph widely studied in the literature  \cite{dall_christensen}.
 To retain the speciation conditions explored by Costa {\it et al.} the parameter $G$ is set so that the ratio $G/B$ remains below 0.2, and the mutation rate is fixed at $0.001$.

From the simulation results, we construct the GSN for both, transient and stationary configurations. The networks consist of a few hundred nodes and connections between every pair of nodes, whose sequences have Hamming distances below the threshold.
 For the definitions of the species,  we use the Biological Species Concept \cite{mayr} according to which two individuals belong to the same species if there exists a chain of compatible individuals that establishes gene flow between them. In the framework of networks,  this means that the nodes representing the individual belong to the same connected component.

Following the same approach  of Costa {\it et al}, we simulate speciation via
computational realizations of the de Aguiar  model \cite{modelo_de_aguiar}.
However, the original model has some algorithmic aspects which attempt   to make it more
realistic, but that represent severe drawbacks for our purposes.  Our adapted
approach allows for a complete control of the parameters and the realization of
a thorough study of the regions of interest. The main  differences from the original version are:

\begin{itemize}
    \item[(a.)] Individuals are not confined to a discrete grid but distribute freely in the available space.
\item[(b.)]         If mating conditions are favorable (namely, if there is a compatible mate in its neighborhood), every mother from one generation gives rise to exactly one offspring, which will take her place in the next generation (when compatibility fails, the offspring is artificially generated from a randomly chosen mother —see above—). This ingredient allows the population size to remain constant, in contrast to the original model, in which individuals can die without giving rise to an offspring (even when compatible mates are available), or can have more than one descendant in their reproductive event (and this is not a case of replacement due to a lack of favorable conditions), resulting in a slight fluctuation of the population size across generations (see \cite{modelo_de_aguiar}).

\item[(c.)] Individuals do not move in the physical space.  We find that this
adaptation allows for a clearer analysis of each speciation scenario at a
significantly lower computational cost and without changing the qualitative
outcome of the original model. Quantitative aspects of the model   such as
transition thresholds and characteristic speciation times do change in the
adapted model, without altering the overall features of the observed phenomena.
\end{itemize}

 We stress that the modifications we introduce to the model do not alter its fundamental outcome.

\section{Simulation Results}\label{sec:simulations}

\begin{figure}
\begin{tabular}{cc}
\begin{tabular}{ccc}
\includegraphics[width=0.3\columnwidth]{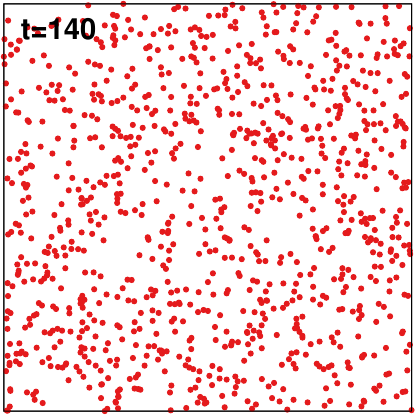} & \includegraphics[width=0.3\columnwidth]{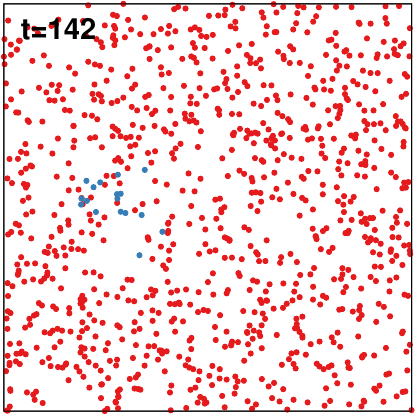} & \includegraphics[width=0.3\columnwidth]{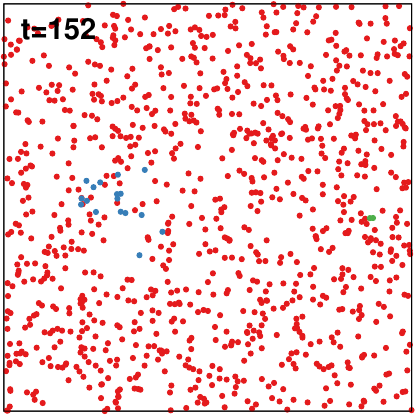}\tabularnewline
\includegraphics[width=0.3\columnwidth]{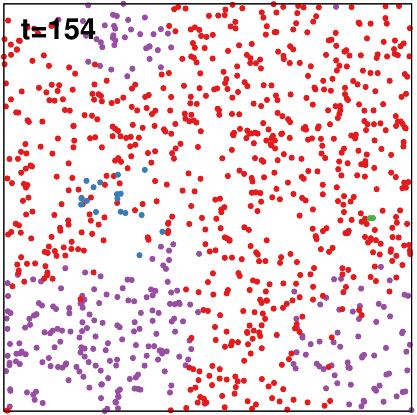} & \includegraphics[width=0.3\columnwidth]{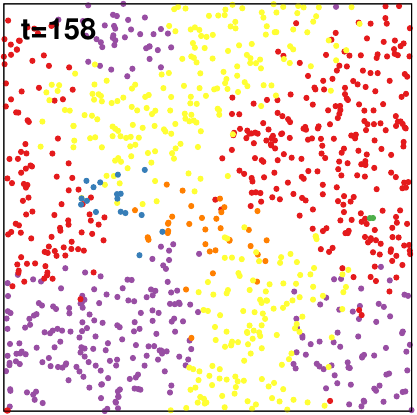} & \includegraphics[width=0.3\columnwidth]{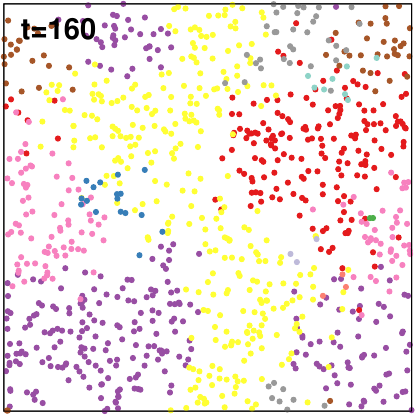}\tabularnewline
\includegraphics[width=0.3\columnwidth]{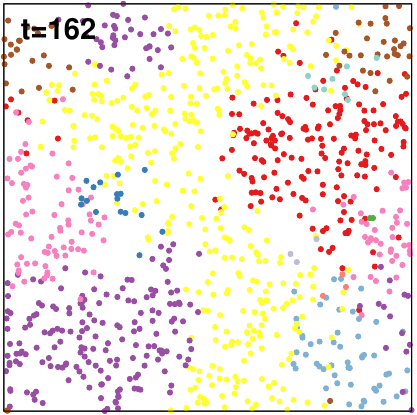} & \includegraphics[width=0.3\columnwidth]{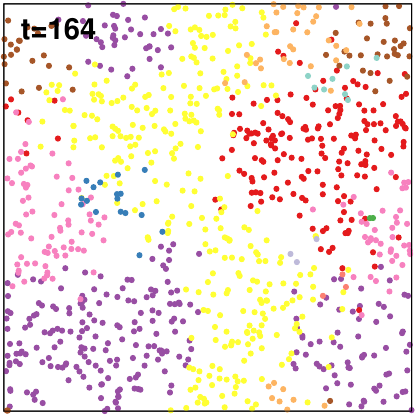} & \includegraphics[width=0.3\columnwidth]{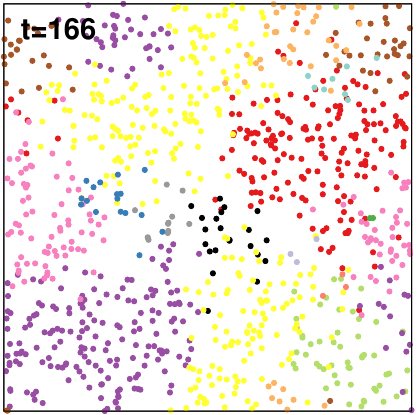}\tabularnewline
\end{tabular} &
\raisebox{-0.265\height}
{\includegraphics[width=0.12\columnwidth]{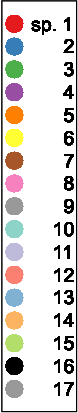}} \tabularnewline
\end{tabular}

\caption{Sequence of snapshots  showing speciation over time for a specific simulation of the model with $S=12$, $B=1280$ and $G=140$. The species are shown in different colors. The  selected times correspond to  speciation  events of two types: splitting (in the form of `fission' or `evaporation') and  merging or extinction  of species (see main text).}
  \label{fig:capturas}
\end{figure}

\begin{figure}[h]
    \centering
\begin{subfigure}[b]{0.98\textwidth}
        \vspace{-2cm}
        \includegraphics[width=0.85\columnwidth]{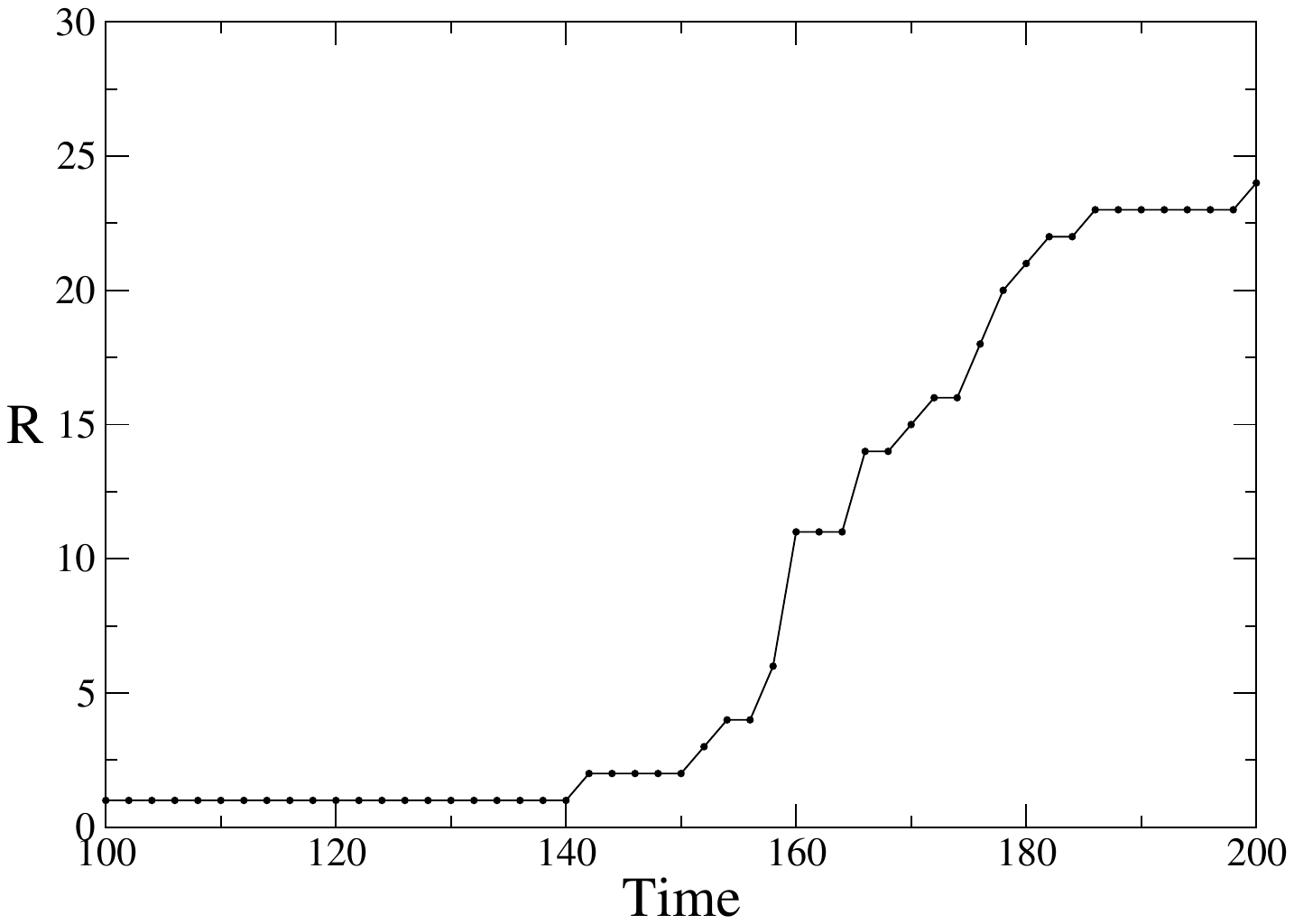}
    \end{subfigure}
    \hfill
    \begin{subfigure}[b]{0.98\textwidth}
        \includegraphics[width=0.8\columnwidth]{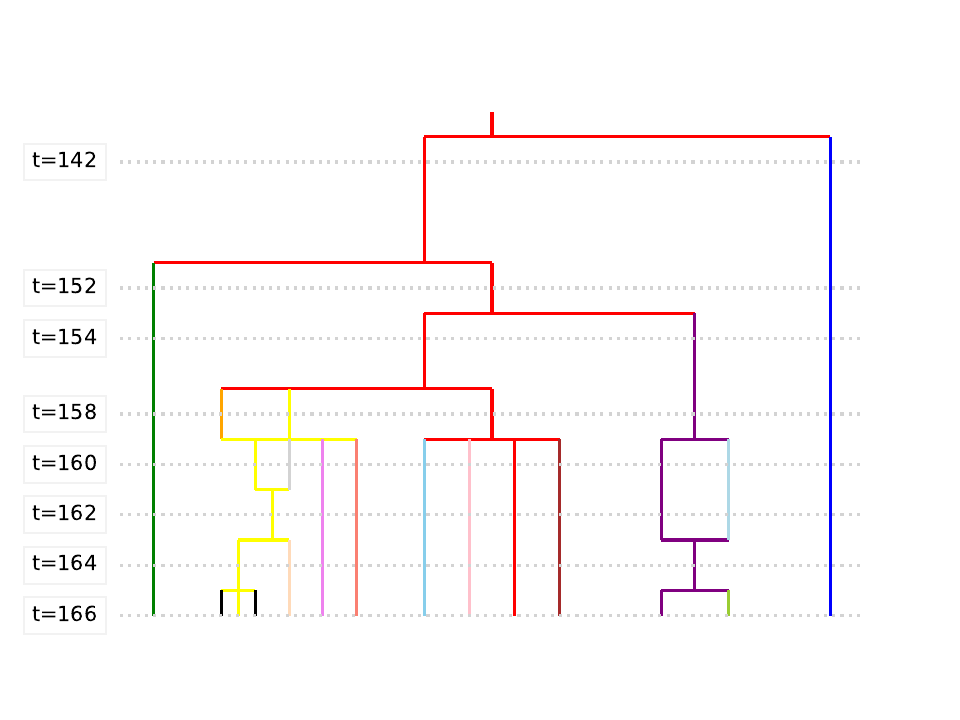}
    \end{subfigure}
    \caption{
Upper panel: number of species as a function of time for a simulation with parameters $B=1280$, $G=140$, and $S=12$.  Lower panel: tree diagram of the speciation and extinction events from $t=140$ to $t=166$ for the same simulation. See also the main text and figure \ref{fig:capturas}.
    }
    \label{fig:richness}
\end{figure}

\begin{figure}[ht]
    \centering
    \begin{minipage}{\textwidth}
        \centering
        \begin{minipage}[t]{0.333\textwidth}
            \includegraphics[width=\linewidth]{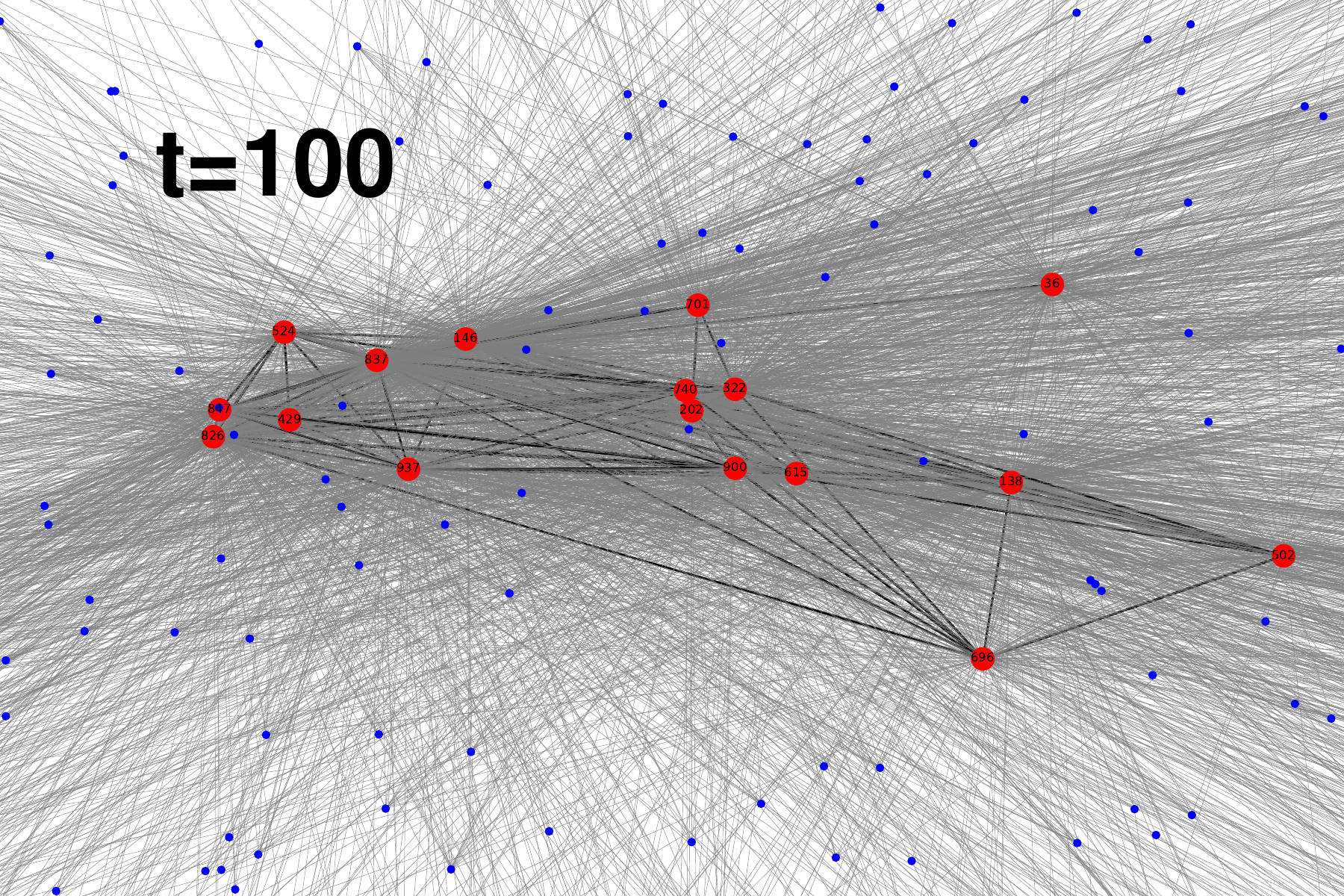}
        \end{minipage}\hfill
        \begin{minipage}[t]{0.333\textwidth}
            \includegraphics[width=\linewidth]{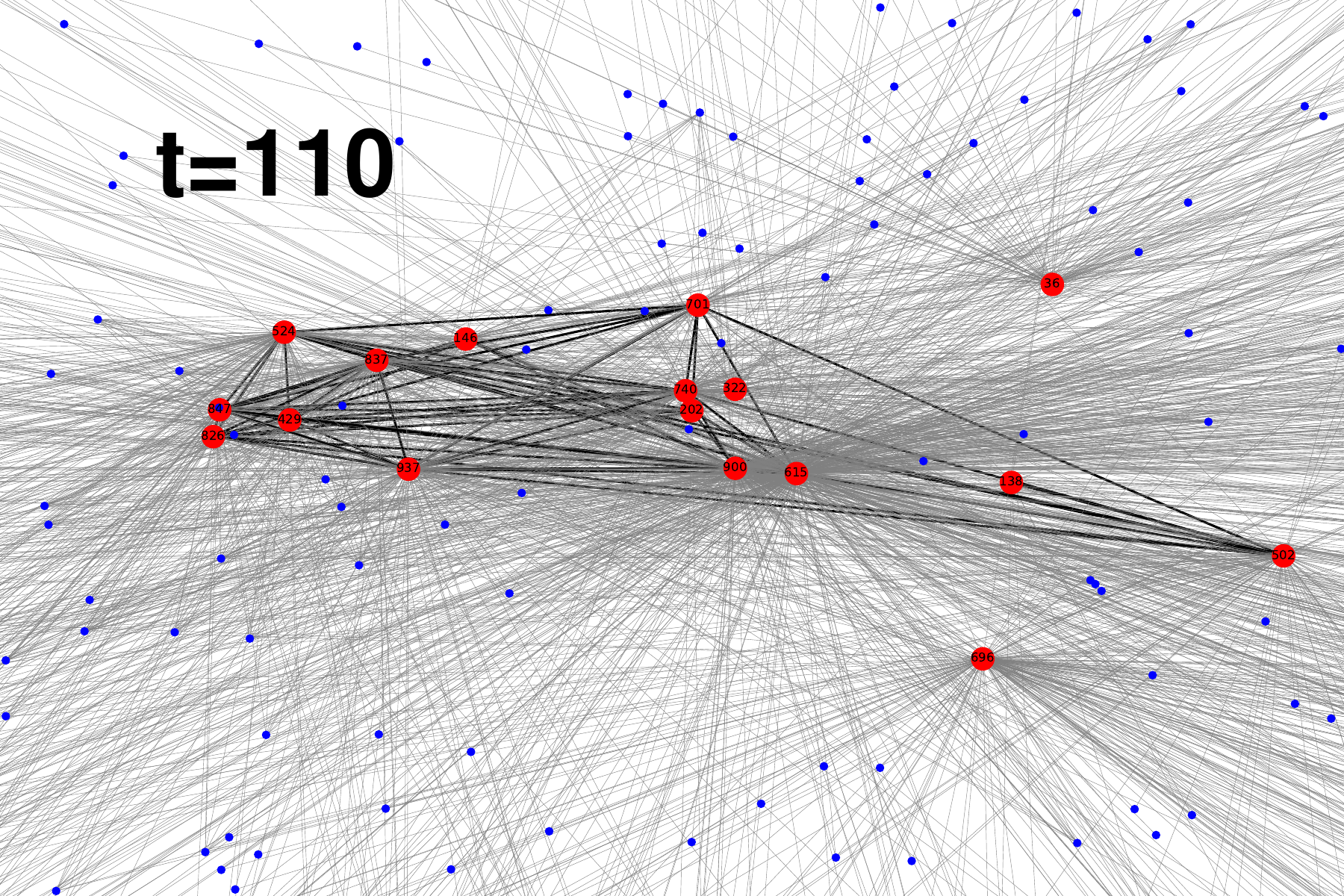}
        \end{minipage}\hfill
        \begin{minipage}[t]{0.333\textwidth}
            \includegraphics[width=\linewidth]{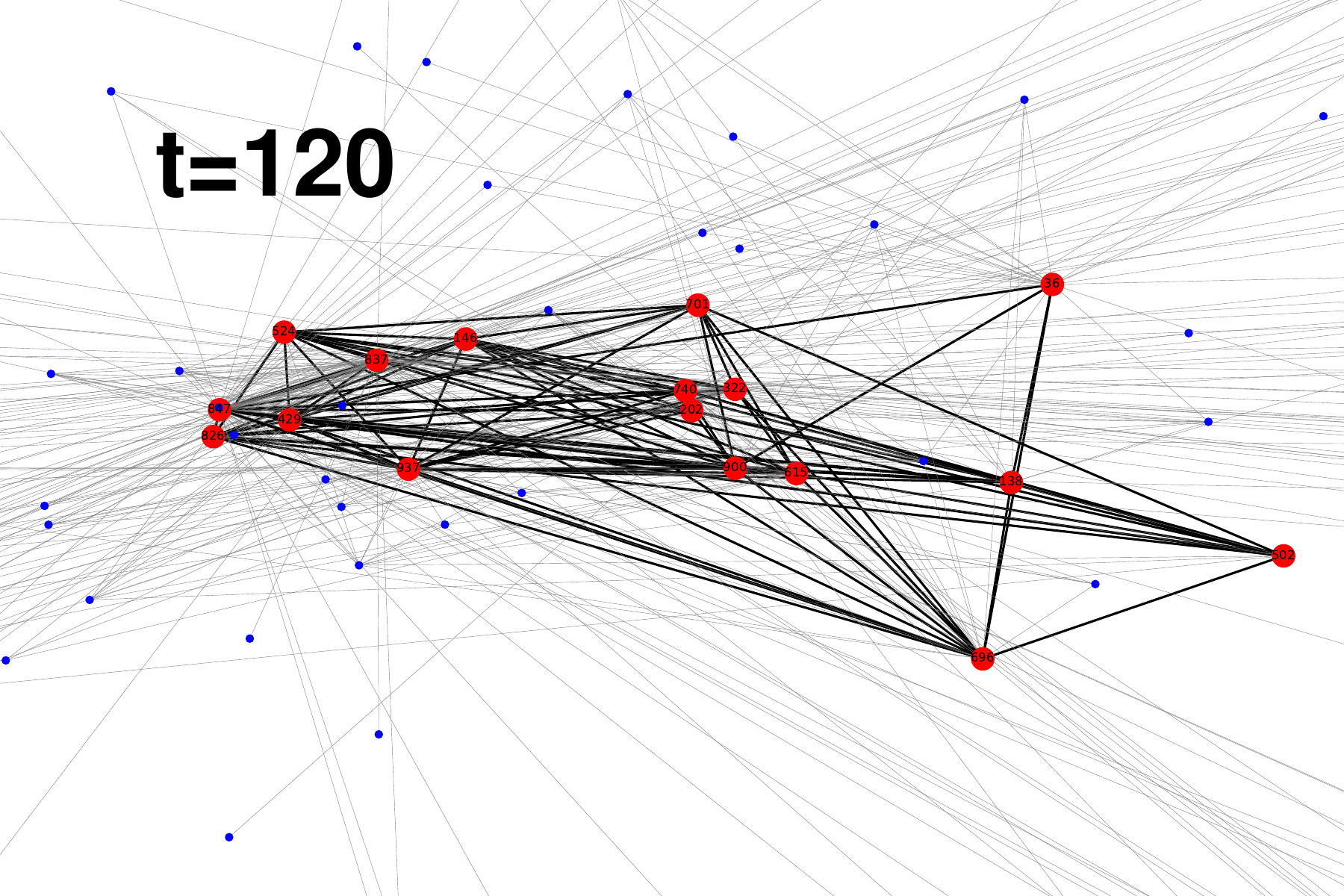}
        \end{minipage}
    \end{minipage}
        \vspace{-0.05cm}
    \begin{minipage}{\textwidth}
        \centering
        \begin{minipage}[t]{0.332\textwidth}
            \includegraphics[width=\linewidth]{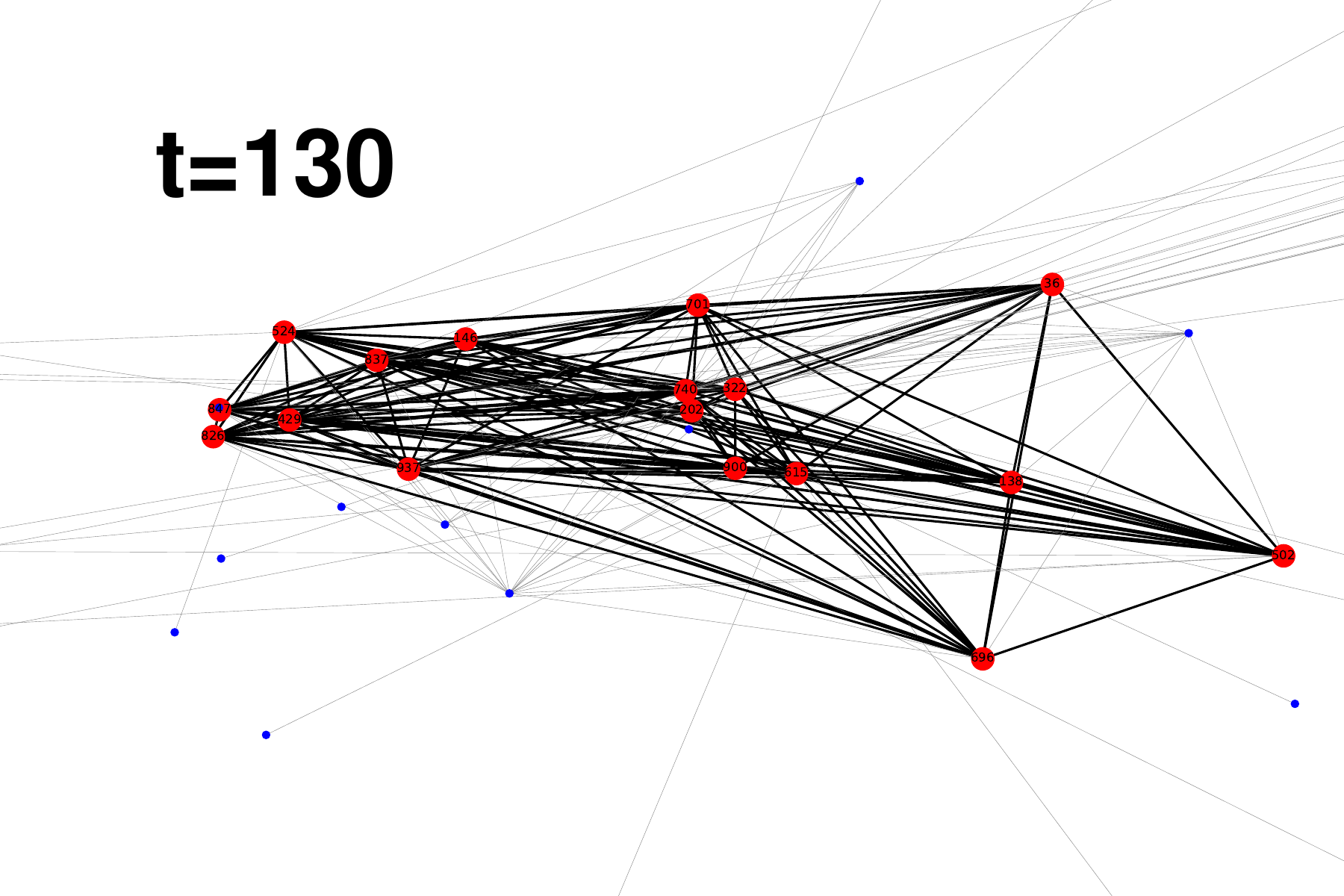}
        \end{minipage}\hfill
        \begin{minipage}[t]{0.332\textwidth}
            \includegraphics[width=\linewidth]{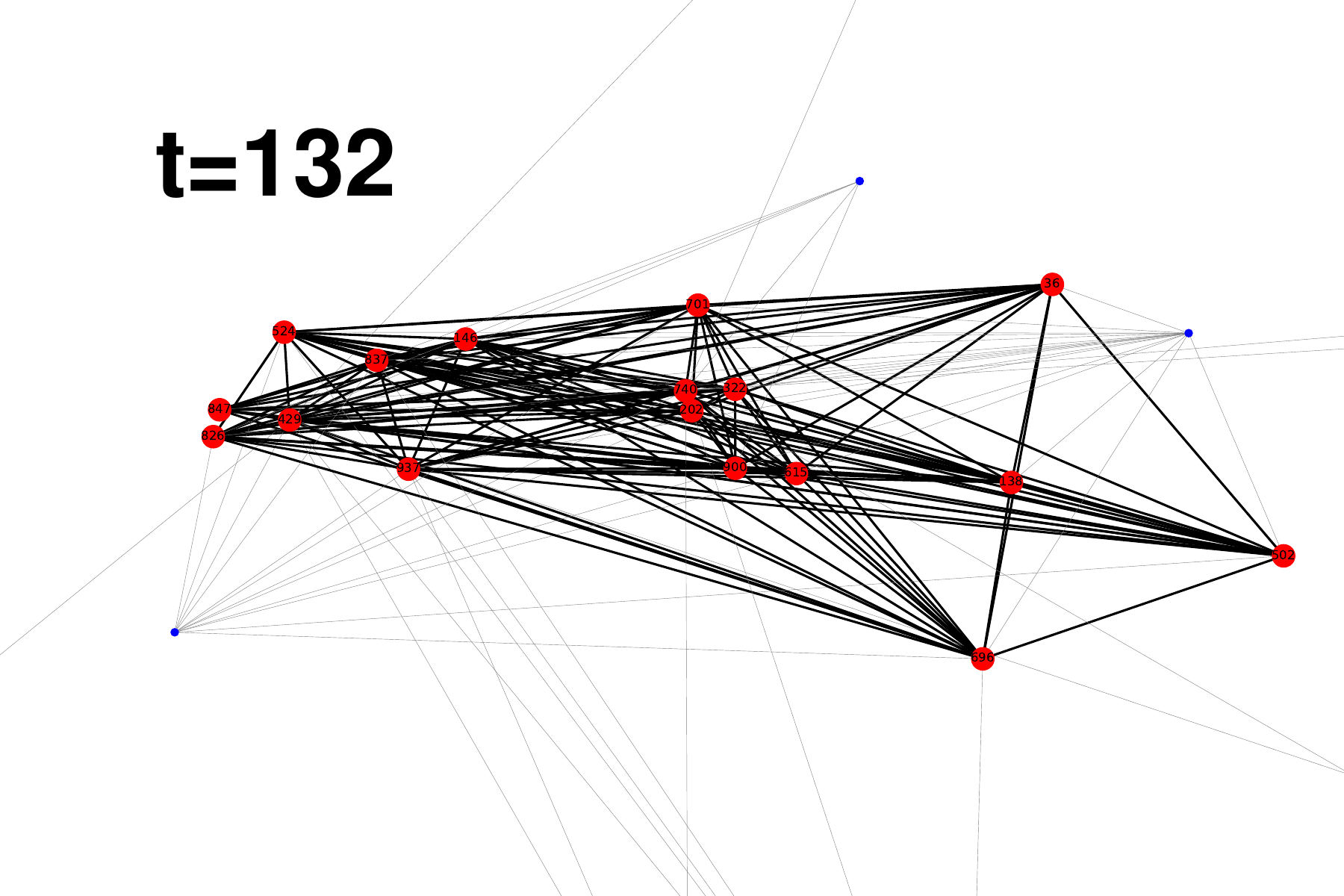}
        \end{minipage}\hfill
        \begin{minipage}[t]{0.332\textwidth}
            \includegraphics[width=\linewidth]{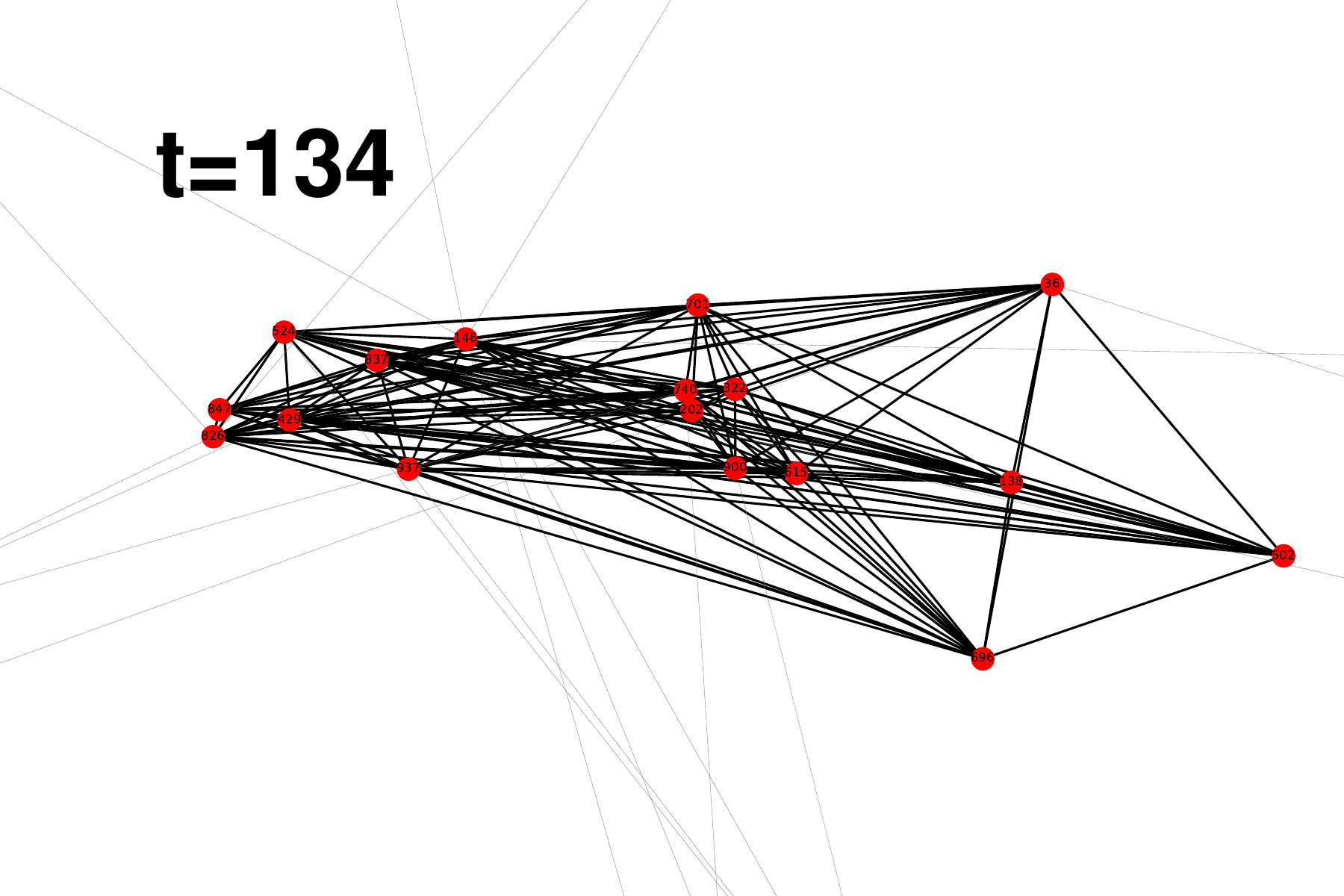}
        \end{minipage}\hfill
   \end{minipage}
    \vspace{-0.05cm}
    \begin{minipage}{\textwidth}
        \centering
        \begin{minipage}[t]{0.333\textwidth}
            \includegraphics[width=\linewidth]{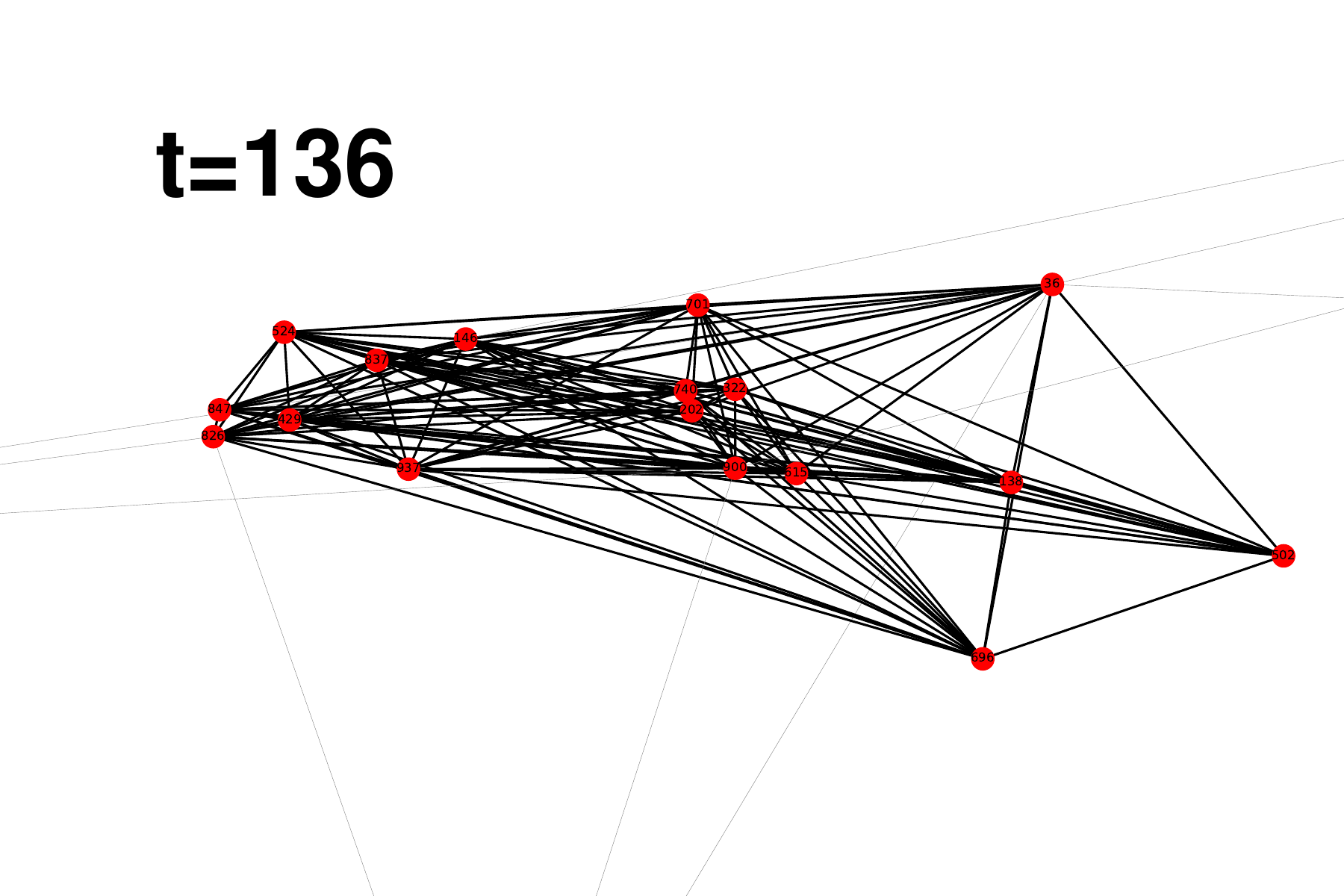}
        \end{minipage}\hfill
        \begin{minipage}[t]{0.333\textwidth}
            \includegraphics[width=\linewidth]{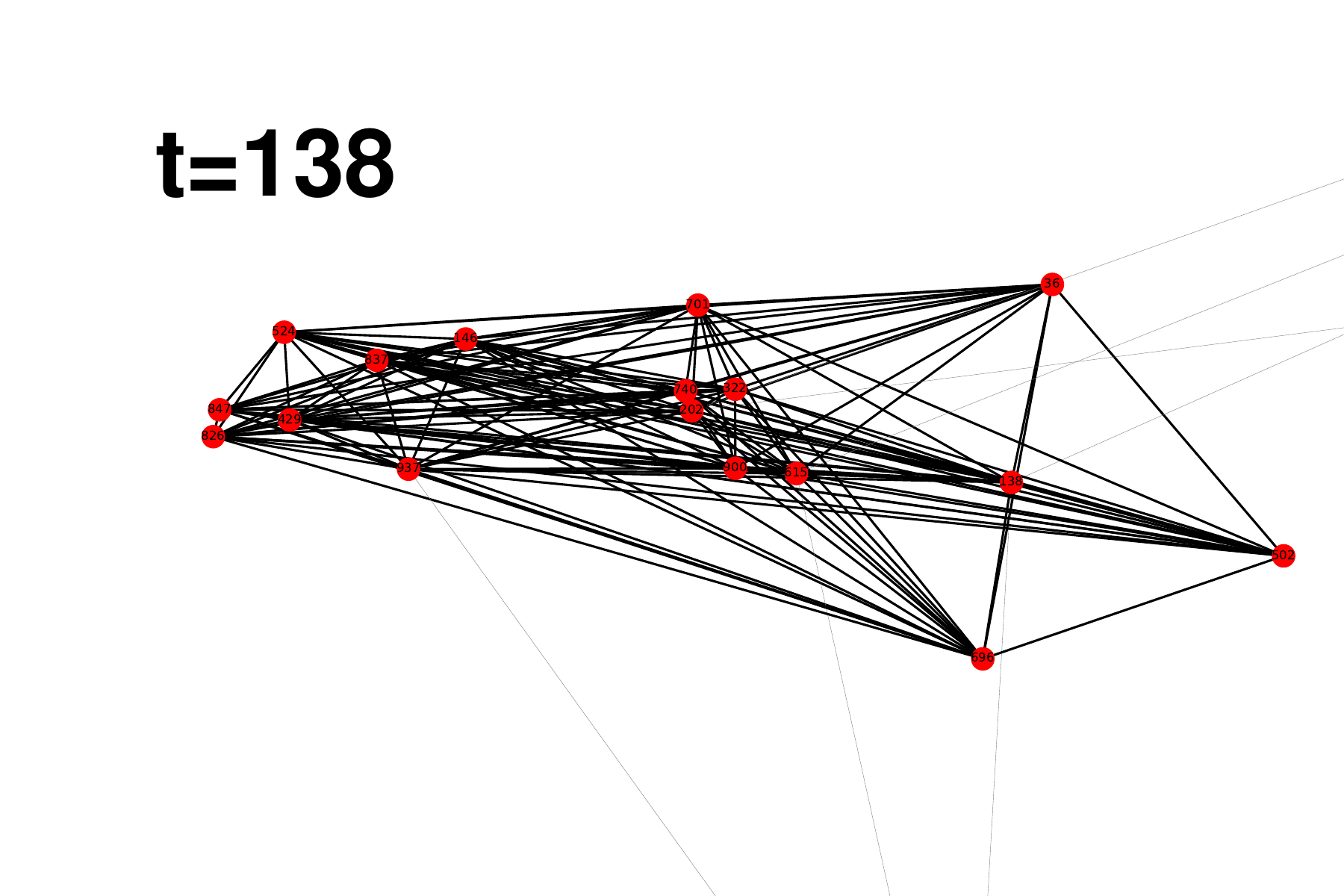}
        \end{minipage}\hfill
        \begin{minipage}[t]{0.333\textwidth}
            \includegraphics[width=\linewidth]{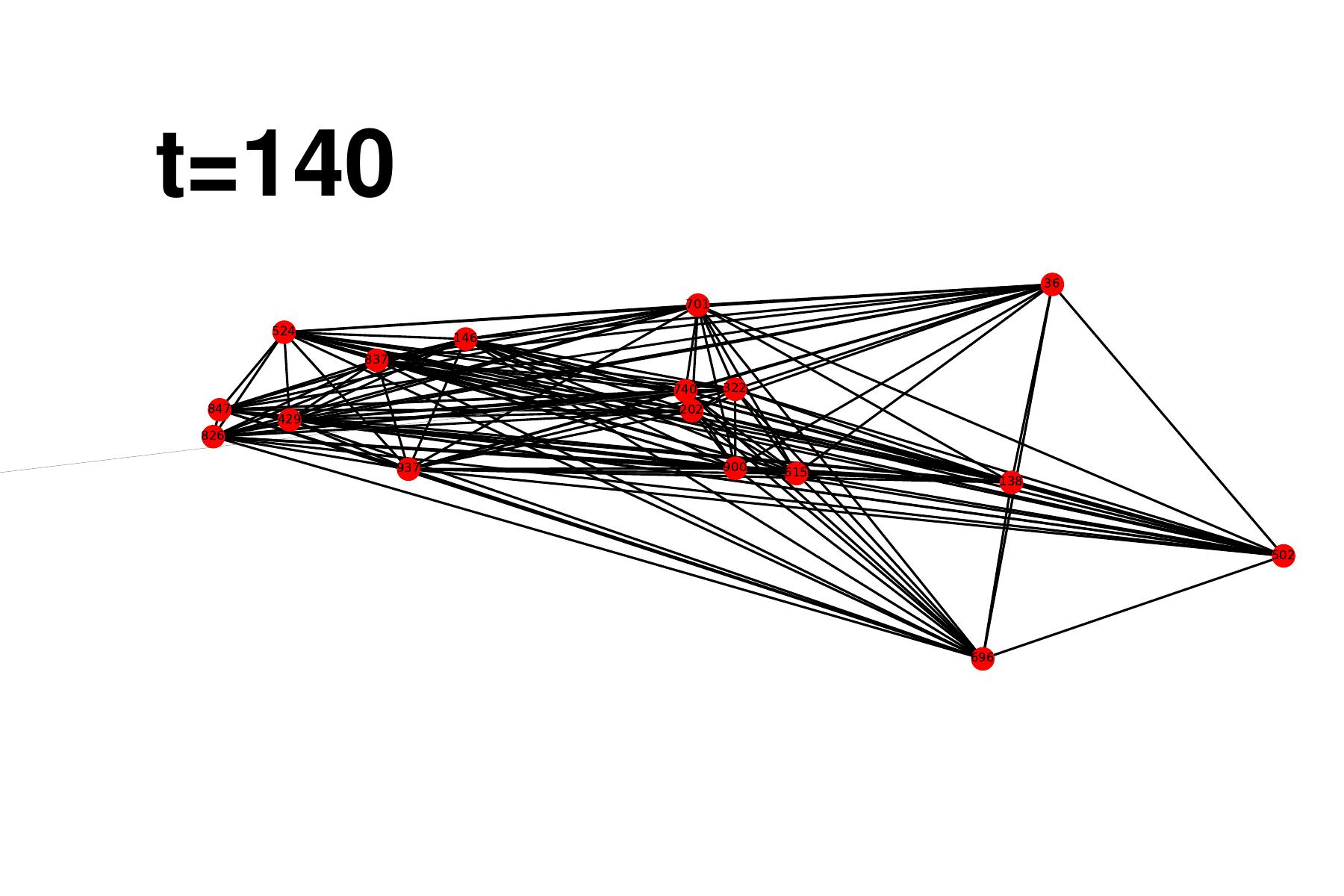}
        \end{minipage}\hfill
       \end{minipage}
    \caption{Snapshots showing the formation of the connected component corresponding to one of the species of the simulation shown  in Figure \ref{fig:capturas}.
We show the species 2, as labeled in the Figure \ref{fig:capturas}. The chosen times correspond to speciation events. See also main text and figures \ref{fig:capturas} and \ref{fig:richness}.}
    \label{fig:red_especie_2}
\end{figure}

The simulations of our adapted de Aguiar model show qualitatively similar behavior to that observed in the context of the original model \cite{costa_1, costa_2, modelo_de_aguiar, diploid_vs_haploid, isolation_by_distance}. Namely, starting from homogeneous populations (with identical sequences, and therefore full connectivity in the associated GSN), after around  $10^2$ generations, clusters of sequences begin to form with a slight spatial overlap. In the context of the biological model, this corresponds to speciation. In the framework of network theory, this shows up as a separation of the network into a set of connected components.
Except for the simulations presented in figure \ref{fig:capturas}, \ref{fig:richness}, and \ref{fig:red_especie_2}, which are strictly descriptive, in all the cases we set the populations size in $N=512$.

In  Figure \ref{fig:capturas} we show snapshots of a typical simulation at times when initial speciation events occur (for this specific simulations, samples are taken at even times, so we are not accounting for the total set of speciation events; however, the purpose of the following paragraphs, summarized in the next tree figures-  is to illustrate qualitatively the main  processes taking place during the course of a simulation, with focus in ramification and fusion of groups of individuals according to its similarity).
   This specific simulation is performed with  $N=1000$ individuals and  a set of parameters of $B=1280$, $G=140$, and $S=12$. The  population  remains as a single species until $t=140$, as observed in  the top-left panel of the figure. The first speciation event occurs at $t=142$, next panel from left to right, where we observe the emergence of a species of 17 individuals splitted  from the original species (species 2, in blue).
At $t=152$, top-right panel, a new species appears (species 3, in green), this time with only two individuals. The temporal line continues in the second  panels strip, where we see at $t=154$  species 1 splitting  into two large groups: one of 702 individuals (which retains the identity of the parent species), and another of 279 (species 4, in violet).
The next event is observed at $t=158$. Species 1  dissociates again, splitting into a group of 356 individuals, which retains the identity of the parent species, another of 311 (species 6, in yellow), and a smaller group of 35 individuals (species 5, in orange, just to the right of species 2).
At $t=160$, we observe the extinction of species 5,  together with the emergence of several new species. Four of them originate from species 1: one with 178 individuals (which retains the parent's species  identity), another with 55 individuals (species 7, in brown), another with 115 individuals (species 8, in pink), and a final one with 8 individuals (species 10, in light blue). Meanwhile, species 6 absorbs the descendants of the extinct species and also undergoes fission itself, producing a core of 300 individuals (which retains the parent species  identity), a species of 40 individuals (species 9, in grey), and two small species of 3 individuals each (species 11, in lila, and species 12, in salmon). These last species are  located in the lower region of the remnant of the primordial species. Species 2, 3, and 4 remain intact from one generation to the next.

These 11 species described previously, undergo few modifications in the snapshot observed two generations later ($t=162$, bottom-left panel). For the first time, species 4 dissociates: of its 279 individuals, only 225 retain their identity, with the remaining 54 becoming species 13 (in dark blue); furthermore, we see the extinction of species 9, whose descendants are absorbed by species 6, which consequently grows to 340 individuals.
Interestingly, both processes are reversed after advancing two more generations ($t=164$, bottom-center panel): we observe the resurgence of the extinct species 9 (whose individuals are  assigned newborn species 14, in light orange),
and the fusion of the 54 individuals of species 13 back into species 4.

Finally, at $t=166$, we observe the resurgence of species 13 (which appears now as species 15, in olive green) and the fission of species 6 into a group of 266 (which retains its identity) and two smaller groups of 22 and 12 individuals, respectively (species 16 and 17, in black and gray, respectively).

The speciation process continues indefinitely, with the number of species stabilizing around 23  by $t \simeq 190$, after which a much slower speciation phase begins (see below). The time course  of this number, known in the literature as ``Richness'', is displayed in figure  \ref{fig:richness}  together  with a tree representation of the speciation events ranging from $t=140$ to $t=166$ (described in detail in the previous  paragraphs).%

As briefly mentioned in the introduction, our goal is to analyze the speciation process using complex network theory, assigning nodes to individuals in the population and links to pairs of individuals whose genetic distance falls below the mating threshold, defined by the parameter $G$. Figure \ref{fig:red_especie_2} shows the region of the network where species 2 forms in the simulation described above. The snapshots correspond to times $t < 142$, i.e., generations prior to the formation of a connected component by this set of 17 nodes (now highlighted in red and identified with numerical labels). From $t = 120$, it is possible to see how the links connecting these nodes to others in the network (in blue) begin to become sparse, while more and more links are added between  the highlighted nodes. At $t = 140$, only a single link remains between this component and the rest of the network, via node with the label 740. Meanwhile, internal links have been established between all existing node pairs, that is, once this group of individuals forms species 2, this species   will constitute a fully connected component.

\subsection{Time evolution of structural properties}
\label{subsec:evolution_struct_props}

\begin{figure}[h!]
\includegraphics[width=0.6\columnwidth]{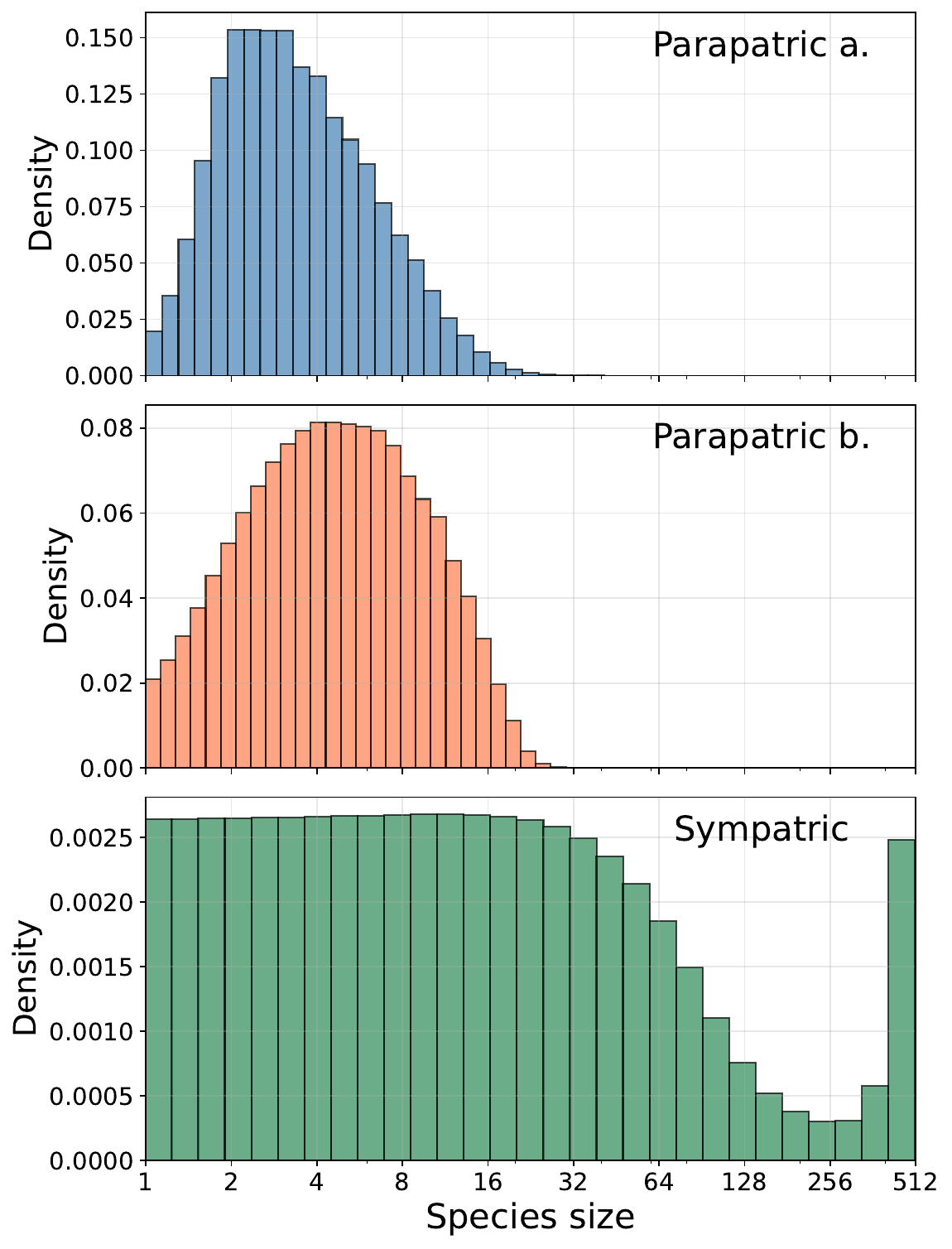}
\caption{Normalized abundances obtained from simulations of the adapted model for  the three parameter  combinations of Table \ref{tab:model_comparison}.\label{fig:abundances}}
\end{figure}

In this section, we study the evolution of the structural properties of the GSN obtained from the simulations. We focus on the three scenarios identified in the work by Costa {\it et al.} through the analysis of phylogenetic trees,  built up using the full version of the model. These three scenarios correspond to particular regions of the $B\!-\!S$ parameter space. Table \ref{tab:model_comparison} shows the typical values of the parameters that characterize each scenario, which in the work of Costa {\it et al.} were determined by quantifying the  tree  balance and branching  acceleration. We consider two configurations for parapatric speciation (high spatial restriction for  mating) and one  for  the sympatric regime. As concluded by Costa {\it et al.}, the size of the genome defines these two characteristics: large genomes favor early speciation and the formation of balanced trees, whereas in the case of small genomes, speciation appears later and results in asymmetric trees. Meanwhile, the sympatric regime is only observed with very large genomes \cite{Aguiar2017}  and gives rise to relatively early speciation, with ``sparsely flowering'' trees lacking  a defined symmetry. The need of very  large genomes for the appearance of the sympatric regime is also observed in the simulations of our adapted model.
 To emphasize the qualitative equivalence of our version of de Aguiar model and the original one, in Figure  \ref{fig:abundances} we show  the abundance curves (number of species of a given size) obtained  from our simulations.
 For comparison, we refer the reader to Figure 3 of the work by de Aguiar \citep{modelo_de_aguiar}, which is also presented in the Supplementary Material of this work, together with a full-phase diagram resulting from our simulations.

\begin{table}[ht]
\centering
\caption{Comparison of speciation scenarios and parameter sets:
the full (original) model vs.\ the adapted model proposed in this work.}
\label{tab:model_comparison}
\setlength{\tabcolsep}{1.5em}
\renewcommand{\arraystretch}{1.2}
\begin{tabular}{lll}
\toprule
\textbf{Scenario} & \textbf{Full model} & \textbf{Adapted model} \\
\midrule
Parapatric a. & \(S = 5,\ B = 150\)        & \(S = 6,\ B = 30\) \\
Parapatric b. & \(S = 5,\ B = 150\,000\)   & \(S = 6,\ B = 1\,200\) \\
Simpatric     & \(S = 40,\ B = 150\,000\)  & \(S = 100,\ B = 10\,000\) \\
\bottomrule
\end{tabular}
\end{table}

The three panels of Figure \ref{fig:abundances} represent the abundances in the parapatric a., parapatric b.  and sympatric scenarios for the parameters of Table \ref{tab:model_comparison} (each -normalized- histogram is the result of  600 realizations of the model, with a population size of 512 individuals and a time of simulation of 3000 generations). It is remarkable  the resemblance  between the shape of the graphs for the parapatric configurations and that  of  panels d. and e. of Figure 3 of reference \cite{modelo_de_aguiar}. Notably, the `log-normal' shape  is not longer evident  in the graph corresponding to the sympatric conditions. However, in the pioneering paper of de Aguiar, sympatric speciation is not investigated, so  there is not a reference curve to compare with. To the best of  our knowledge, this the first attempt to obtain an abundance curve in sympatry from simulations, so it would be of significant interest to analyze its statistical properties in the future.
As a preliminary insight, we recognize an increase of the abundance at large values of species sizes which is consistent with the `evaporation phase'  dominating almost the whole range of speciation regime in the sympatric scenario  (see below). This is very interesting, because it establishes  again a measure which can be added to the arsenal of tools for distinguishing speciation modes.

In the rest of our work, we aim to complement the phylogenetic tree characterization of Coste {\it et al.} by studying structural properties of the networks representing populations and their genetic similarity ties. For that purpose, we simulate populations of $N=512$ individuals  (with initially identical sequences, and therefore full connectivity in the
associated GSN), in the parameter regions corresponding to each speciation scenario and build up the corresponding GSN.

\begin{figure}[h!]
    \centering
        \includegraphics[width=0.8\columnwidth]{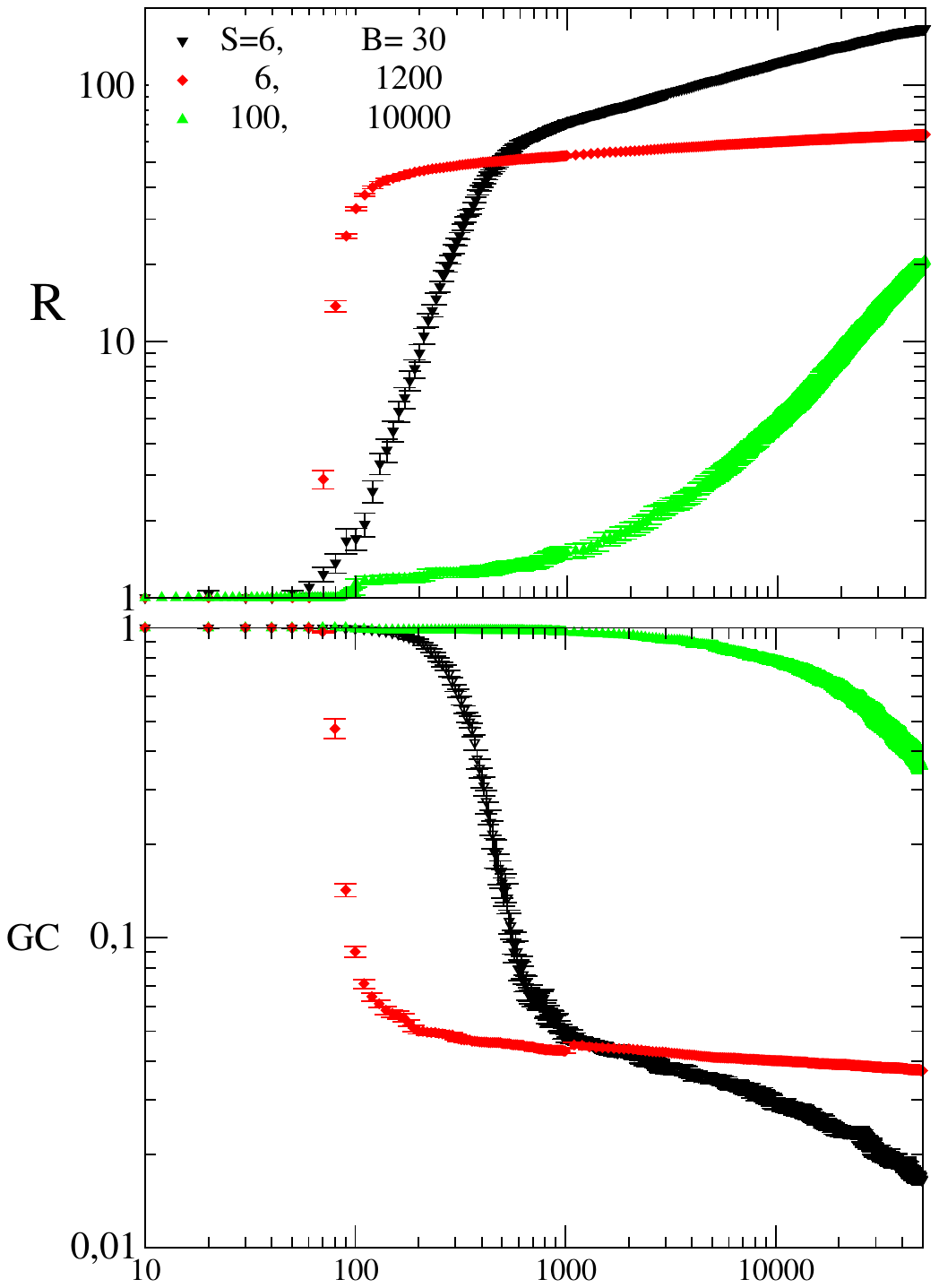}
    \caption{Simulations for the Parapatric a. ($S=6$, $B=30$) , Parapatric b. ($S=6$, $B=1200$), and Sympatric ($S=100$, $B=10000$) scenarios. The upper panel displays richness $R$, and the lower  the fraction of nodes belonging to the giant component $GC$. In Supplementary Material it can be found the outcomes for longer simulation times.}
    \label{fig:long_times}
\end{figure}

Before presenting the evolution of the different structural properties of the networks, it is worth describing the long-time behavior of the number of formed clusters (number of species) and the giant component (predominant species size), because  these metrics  determine the existence of different phases of speciation. The  results presented in  Figure  \ref{fig:long_times}  correspond to averages performed over 30 realizations in the parapatric cases, and 100 realizations in the sympatric one.

Both set of curves of  Figure \ref{fig:long_times} show that, even when the richness and the giant component  exhibit different behaviors in each scenario, three phases can be clearly distinguished in all the cases.
 In all scenarios there exists a short-lasting initial phase in which individuals  still represent a single species ($R=1$, $S=1$), in spite of the fact that they have already started to differentiate. This phase is followed by a another one of variable duration, in which we observe very steep slopes in both parapatric scenarios, and
a much smaller  slope in the sympatric case. This phase spans between  $10^2$ to $10^3$ generations.
Finally, for times on the order of $10^4$ generations or longer, we see a reversal of the previous behavior: both, richness and the giant component, exhibit little variation in the parapatric configurations, while there appears to be an acceleration in the sympatric case.
 The state at infinitely large times appears to be that of maximum entropy, for all the cases. It is remarkable that   for both, the richness and the giant component, the curves displayed in Fig. \ref{fig:long_times} are consistent  with the dendograms  obtained by Costa {\it et al.} in Refs. \cite{costa_1, costa_2}. Speciation
takes place following approximately symmetric fragmentation patterns in the parapatric scenarios (a.) and (b.). In the sympatric case, however, a large central species continually releases small clusters.
Borrowing the vocabulary from nuclear physics, we will call these two speciation patterns `fission' and 'evaporation' patterns of speciation, respectively.

In the following paragraphs, we describe the evolution of various structural properties of the GSN as a function of time.
For each specific value of the parameters $S$ and $B$ (with $G=0.1 B$), we perform $30$ realizations in the parapatric cases, and $100$ realizations in the sympatric case. We set simulation times to  $2\times 10^3$ generations for parapatric a., $ 10^3$ for parapatric b., and $ 10^4$ for sympatric.
As in the simulation described in the previous section, we set  $L=100$, and $\mu=0.001$ in all cases. Samples are taken every 10 generations.

\begin{figure}[h!]
\includegraphics[width=0.9\columnwidth]{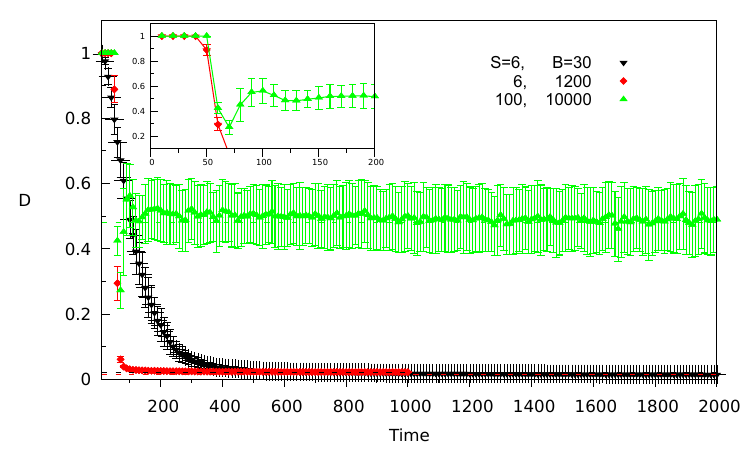}
 \caption{\label{fig:conn_vs_time} Normalized mean degree or network connectivity  as a function of time. The corresponding post-transient values  are shown with dashed lines. The inset shows a zoom of the region of the onset of speciation. For clarity, parapatric scenario a. has been excluded.}
 \end{figure}

In Figure \ref{fig:conn_vs_time}  the evolution of  network connectivity is shown. This metric is  defined as the fraction of linked node pairs  relative to the total number of node pairs within the network, and it also represents the (normalized) mean  degree of the network ($D$). In Appendix A, we provide the formal definition of the properties calculated in this section.
At the beginning of the simulations, connectivities are close to one, which is consistent with a fully connected topology, as expected for identical genomes. Then, a  sharp drop is observed for the three simulated scenarios, with an post-transient value close to zero for the parapatric cases.  The sympatric case, instead, remains  well above zero until the end of the simulation.
Furthermore, a pattern of strongly damped, yet very clear  oscillations can be observed at the onset  of the speciation phase, for the sympatric scenario.

\begin{figure}[h!]
 \includegraphics[width=0.9\columnwidth]{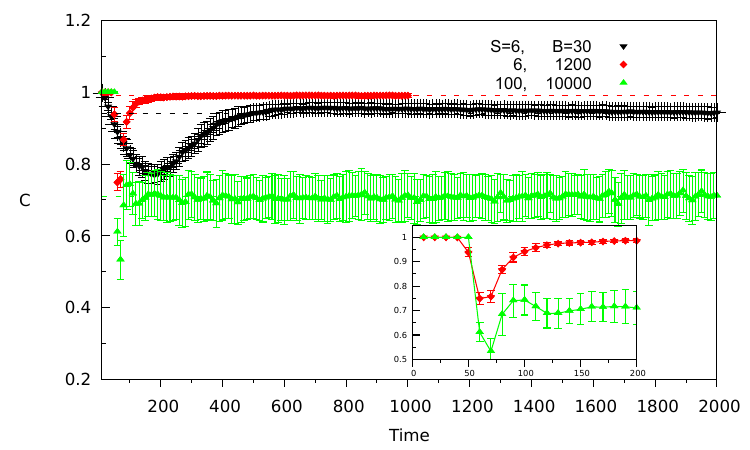}
 \caption{\label{fig:clus_vs_time} Mean clustering coefficient as a function of time. The corresponding post-transient values are shown with a dashed line. The inset shows a zoom of the earliest speciation phase of the simulations for scenarios (a.) and (c.)}
 \end{figure}

Figure \ref{fig:clus_vs_time} shows the evolution of the mean clustering coefficient ($C$). This coefficient is obtained by averaging  the local clustering coefficient over the whole network. This is defined as the number of actual  connections among the neighbors of a given node relative to the total number of potential connections between them.  That is, $k(k-1)/2$, where $k$ is the degree of the node.
 Again, a sharp decrease is observed at the initial events of speciation, after a short phase of high local clustering. However, in  this case the drop down is rapidly followed by a recovery to the initial values for the parapatric cases, and to a value close to $0.7$  in the sympatric case. This is shown  in inset of Fig. \ref{fig:clus_vs_time}. The post-transient  values appear to depend entirely on the parameter $S$. Sympatric oscillations show  a similar pattern to that of Figure \ref{fig:conn_vs_time}.

\begin{figure}[h!]
 \includegraphics[width=0.9\columnwidth]{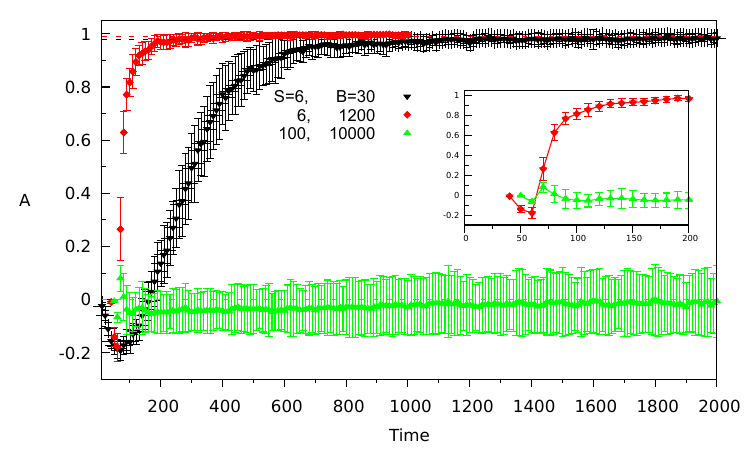}
 \caption{\label{fig:ass_vs_time} Degree   assortativity as a function of time. The corresponding post-transient values are shown with  dashed lines. The inset shows a zoom of the earliest speciation phase of the simulations for scenarios a. and c.}
 \end{figure}

Figure~\ref{fig:ass_vs_time} shows the evolution of the degree assortativity as a function of time. This coefficient, bounded between $-1$ and $1$, represents the correlation between nodes according to their degree $k$.
For this metric we observe a different behavior.  Starting from zero, all the   curves  immediately become negative. In the course of the fast phase  of  speciation (fission pattern), the parapatric curves  rapidly increase towards one (high assortativity), while the sympatric curve slowly approaches to zero, with  increasingly large fluctuations. Sympatric oscillations, on the other hand, appear to be almost negligible.

\begin{figure}
 \includegraphics[width=0.9\columnwidth]{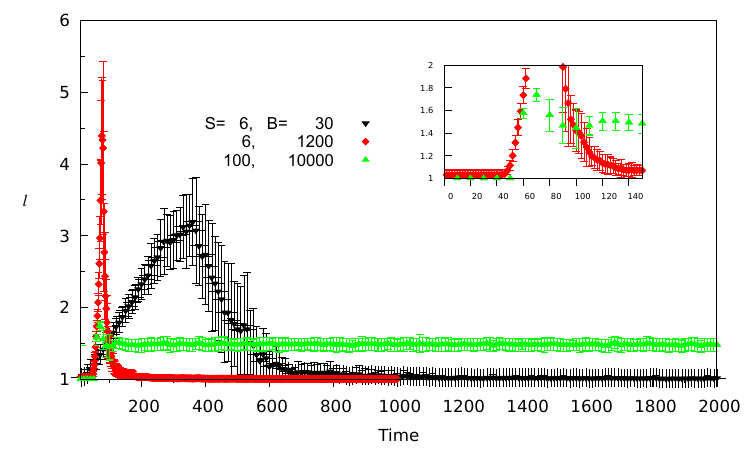}
 \caption{\label{fig:shPh_vs_time} Mean geodesic path as a function of time.  The inset shows a zoom of the earliest speciation phase of the simulations for scenarios (a.) and (c.)}
 \end{figure}

We conclude this section by showing the time course of the average geodesic path  in Fig. \ref{fig:shPh_vs_time}. This metric corresponds to the  shortest path between a pair of nodes, averaged over all pairs of nodes for which a path exists   (pairs with one node in one connected component and the second node in an other connected component, have not a defined geodesic path).
 At the beginning of the simulation, mean geodesic paths are equal to $1$ in all cases. This is consistent with the fact that initially all the individuals are identical (fully connected GSN). As individuals begin to differentiate, the geodesic path starts to increase, specially in the parapatric cases, reaching a maximum at the beginning of speciation. In this region, richness curves acquire their maximum slopes. After this, the geodesic path stabilizes at lower values. The post-transient   values for the parapatric curves suggest that most daughter-connected components are fully connected, contrary to the sympatric case, for which mean geodesic paths approach the value  $1.5$. However, it must be recalled that this phase of sympatric speciation does not persist indefinitely. The long-time behavior would deserve further investigations.

 \subsection{Post-transient properties}\label{subsec:asympt}

  Except for the case of the richness and the giant component, all structural properties show a stabilization after the fission phase, which in the parapatric frameworks finishes at times smaller than $10^3$ generations. Accordingly, we present  the results of the structural metrics calculated at $t=2000$.   The  mean values and standard deviations are obtained taking 100 realizations. The initial conditions and model parameters are the same as that of the simulations presented in the previous section ($N=512$, $\mu=10^{-3}$, $L=100$), except for $S$ and $B$.
 We study the whole range of the parameter $S$, and genome sizes  up to $B=10^5$, for all the cases.

 \begin{figure}[h!]
    \centering
    \makebox[\textwidth][c]{%
    \begin{subfigure}[b]{0.528\textwidth}
   \includegraphics[width=\linewidth]{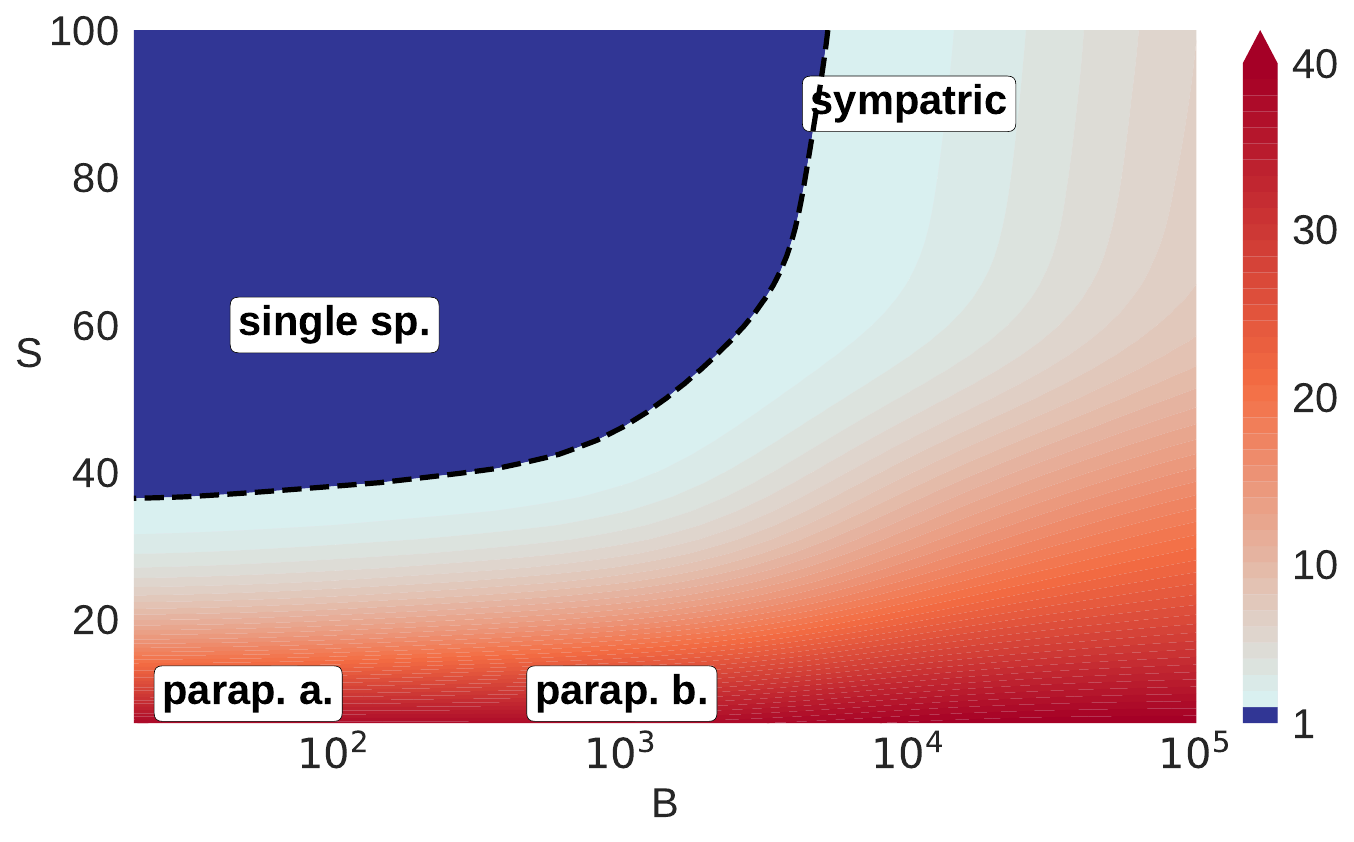}
    \end{subfigure}%
    \hspace{0.5em}\begin{subfigure}[b]{0.624\textwidth}
        \includegraphics[width=\linewidth]{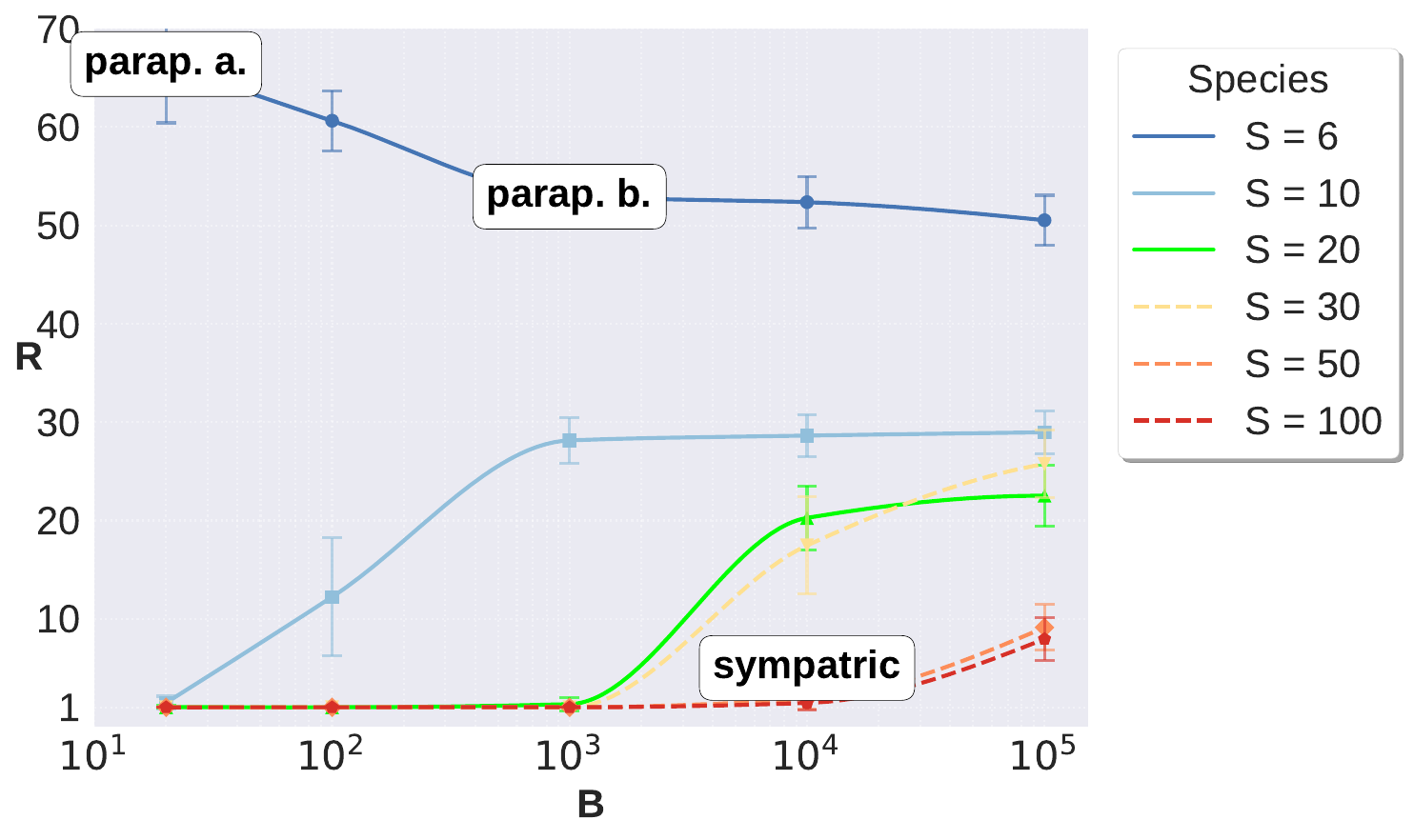}
   \end{subfigure}}
    \caption{Long-time behavior of the richness $R$ as a function of the parameters $S$ and $B$.  The left panel shows a color plot covering the parameter space, in which a dashed line (defined as the contour corresponding to $R=1.5$) separates the single species region  (``single sp.'') from the speciation regions. The right panel presents $R$ vs. $B$ for some specific values of the parameter  $S$.    }
    \label{fig:rich_asymp}
\end{figure}

\begin{figure}[h!]
   \centering
    \makebox[\textwidth][c]{%
    \begin{subfigure}[b]{0.528\textwidth}
   \includegraphics[width=\linewidth]{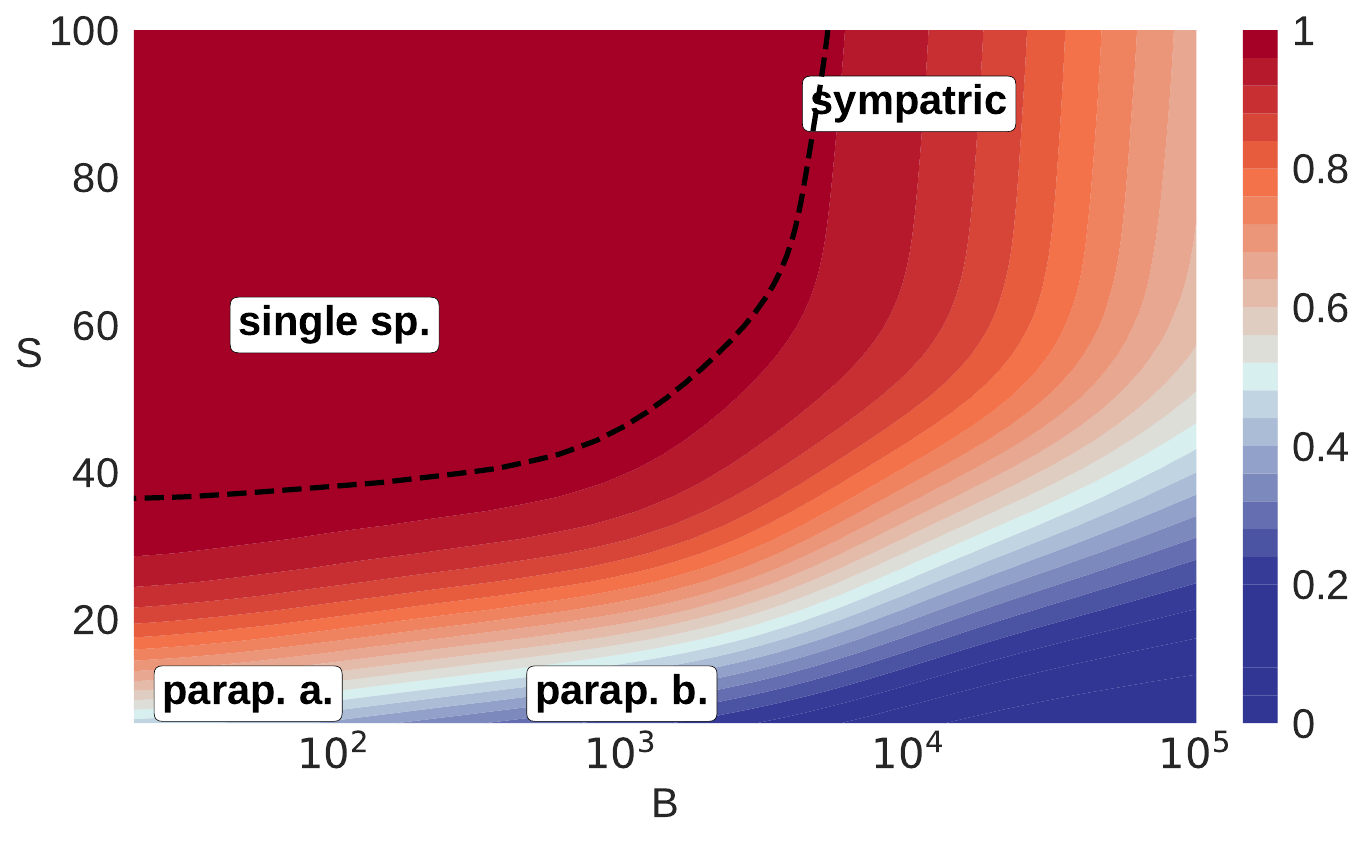}
    \end{subfigure}%
    \hspace{0.5em}\begin{subfigure}[b]{0.624\textwidth}
        \includegraphics[width=\linewidth]{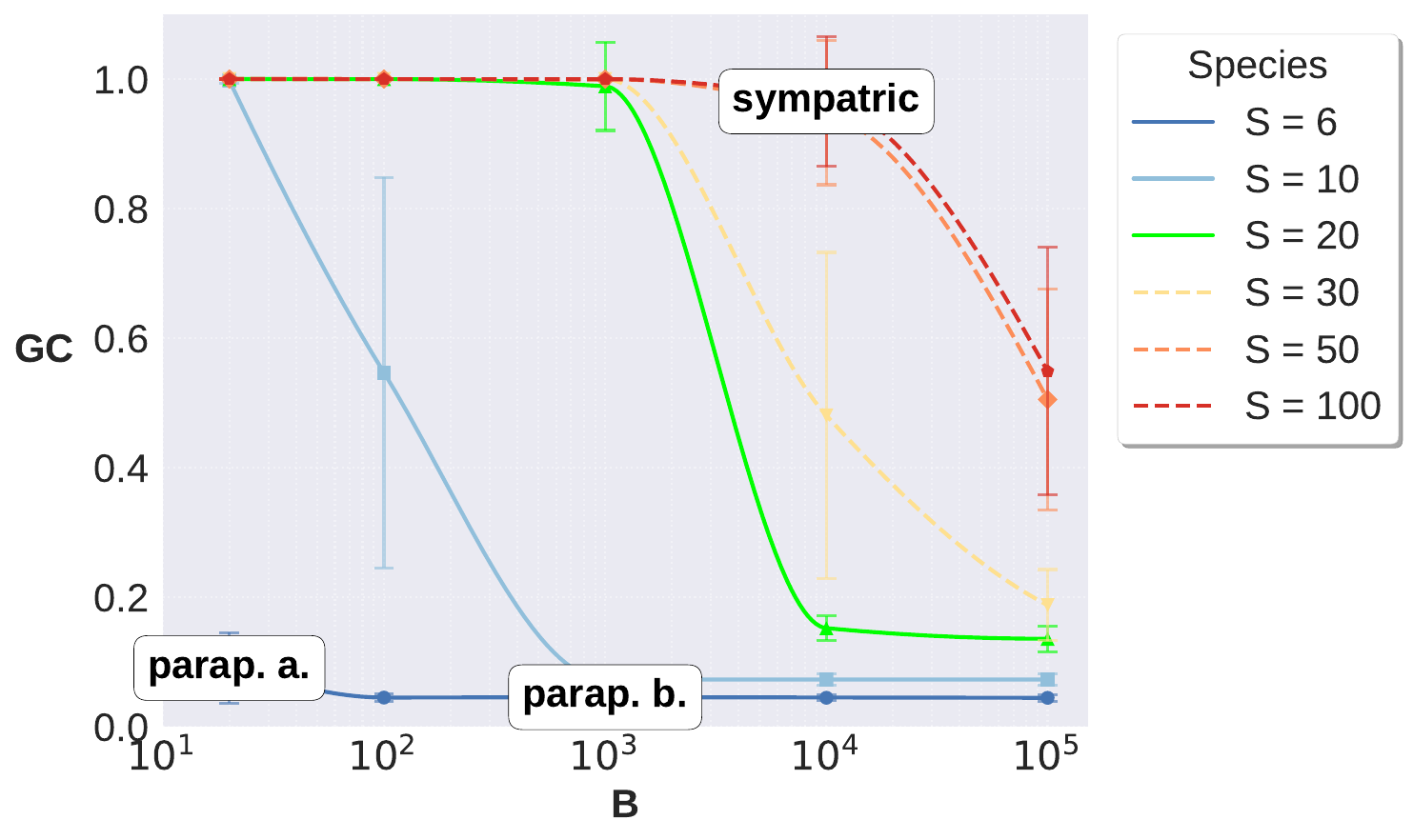}
   \end{subfigure}}
    \caption{Long-time behavior of the giant component $GC$ as a function of the parameters $S$ and $B$.  The left panel shows a color plot covering the parameter space, in which a  dashed line (defined as the contour corresponding to $R=1.5$) separates the single species region  (``single sp.'') from the speciation regions. The right panel presents $GC$ vs. $B$ for some specific values of the parameter  $S$.}
    \label{fig:giant_asymp}
\end{figure}

\begin{figure}[h!]
   \centering
    \makebox[\textwidth][c]{%
    \begin{subfigure}[b]{0.528\textwidth}
   \includegraphics[width=\linewidth]{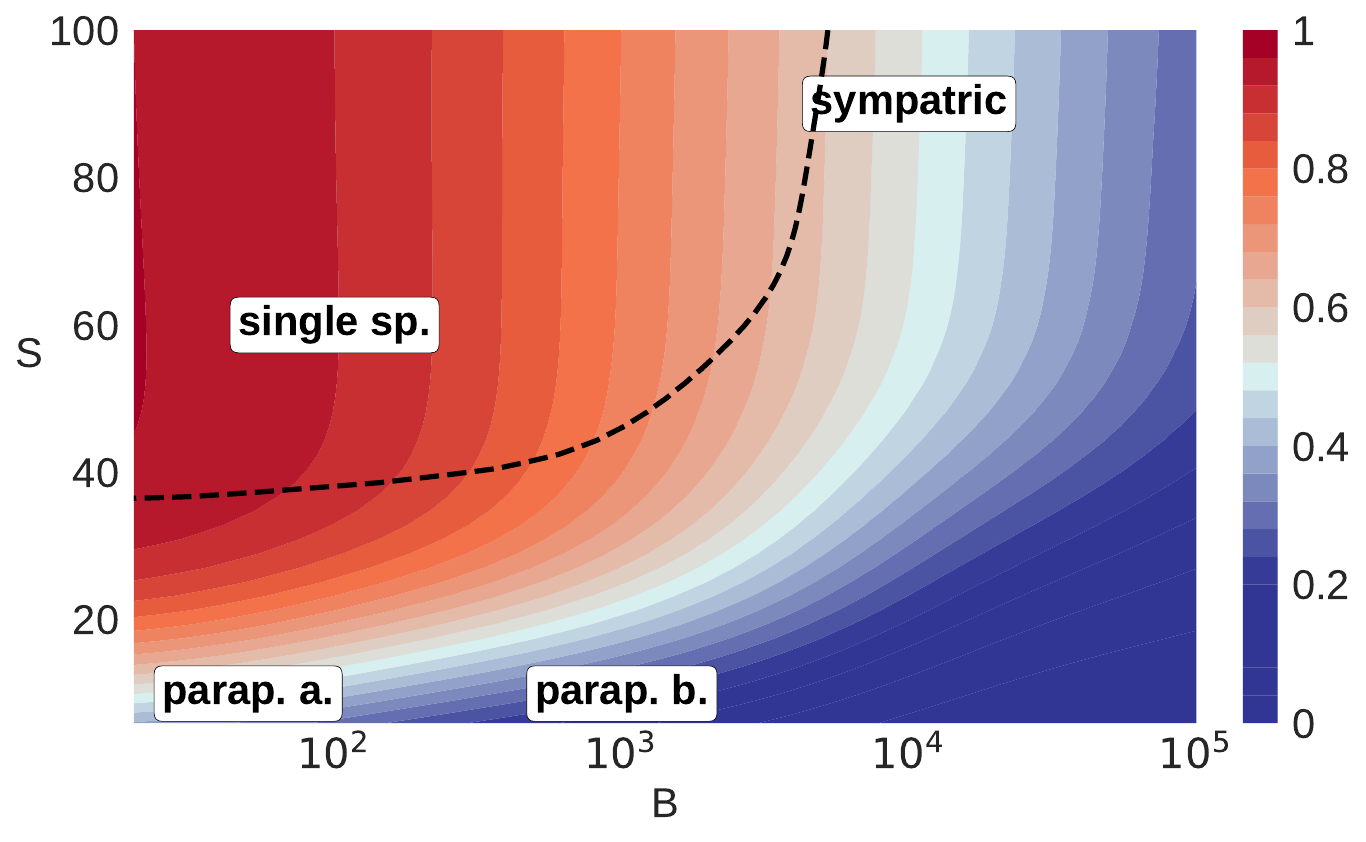}
    \end{subfigure}%
    \hspace{0.5em}\begin{subfigure}[b]{0.624\textwidth}
        \includegraphics[width=\linewidth]{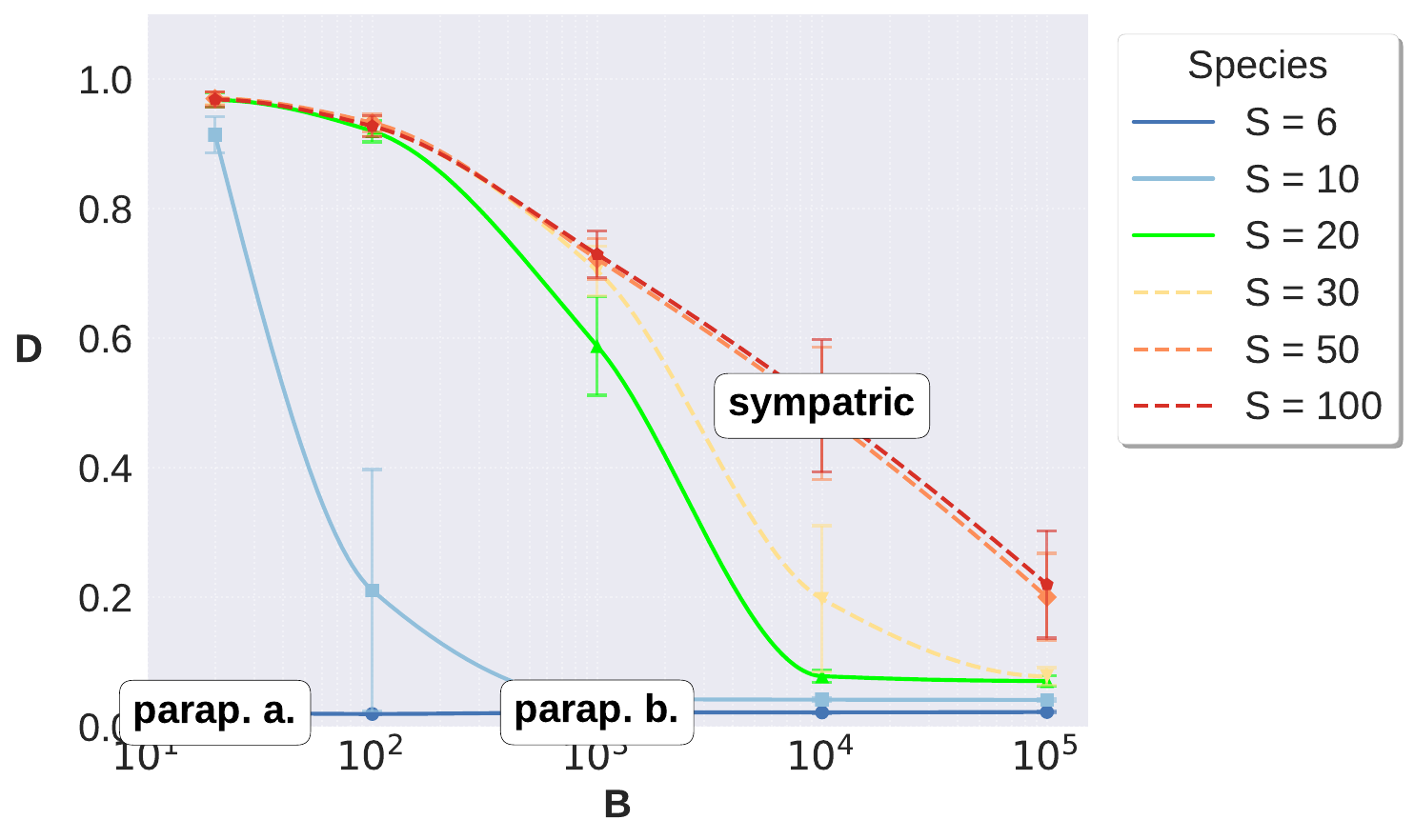}
   \end{subfigure}}
    \caption{Long-time behavior of the mean degree $D$ as a function of the parameters $S$ and $B$. The left panel shows a color plot covering the parameter space, in which a dashed line (defined as the contour corresponding to $R=1.5$) separates the single species region  (``single sp.'') from the speciation regions. The right panel presents $D$ vs. $B$ for some specific values of the parameter  $S$.    }
   \label{fig:conn_asymp}
\end{figure}

\begin{figure}[h!]
   \centering
    \makebox[\textwidth][c]{%
    \begin{subfigure}[b]{0.528\textwidth}
   \includegraphics[width=\linewidth]{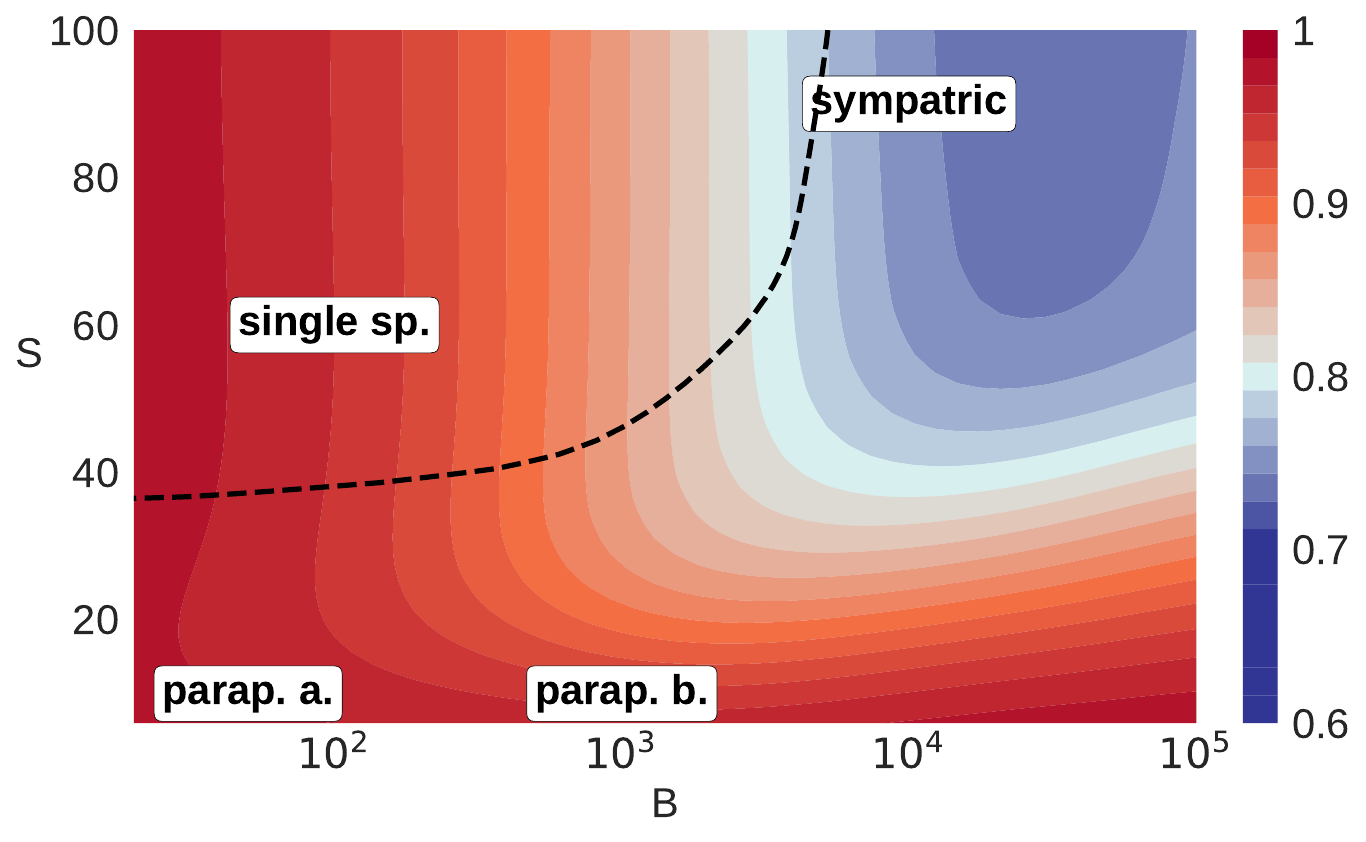}
    \end{subfigure}%
    \hspace{0.5em}\begin{subfigure}[b]{0.624\textwidth}
        \includegraphics[width=\linewidth]{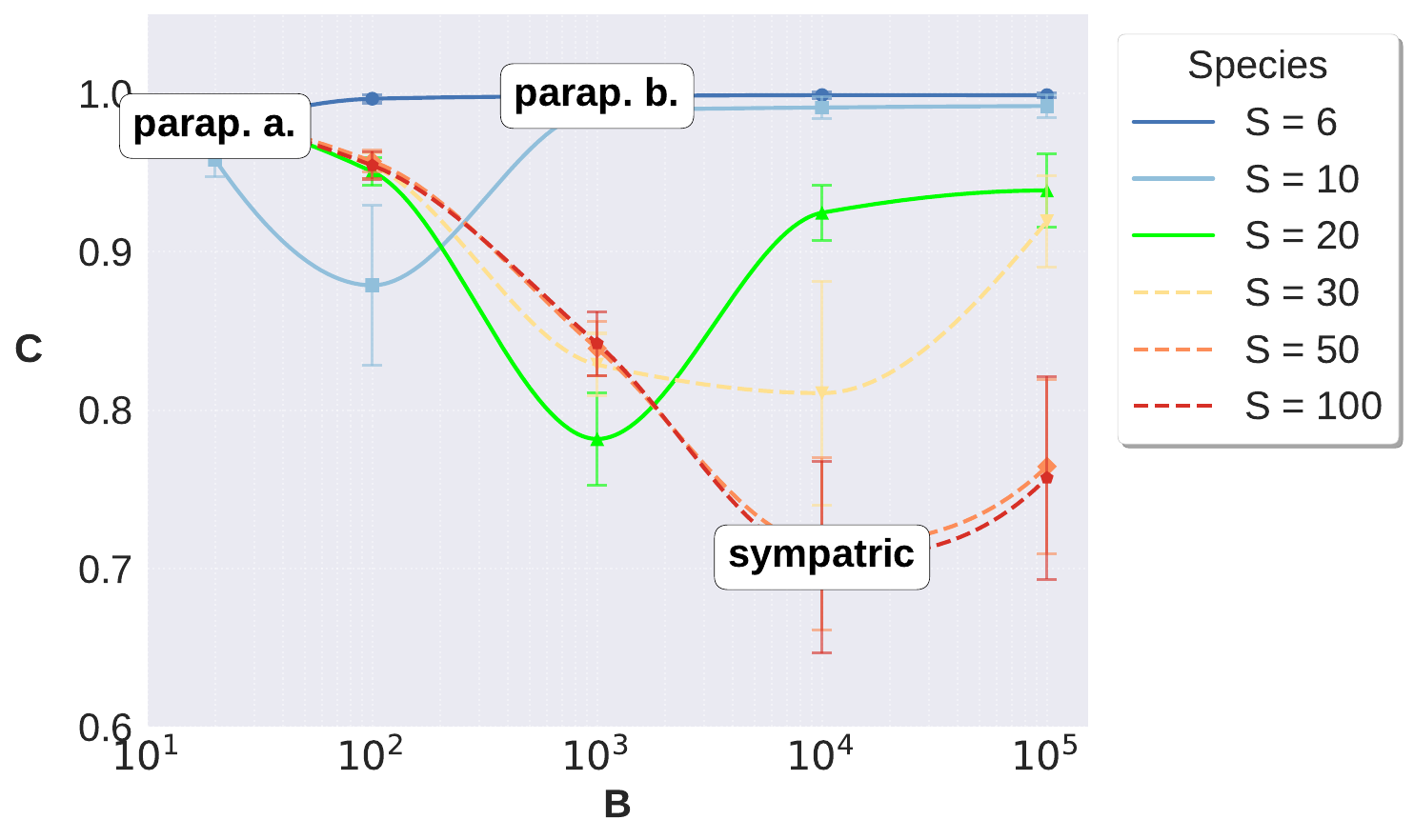}
   \end{subfigure}}
  \caption{Mean local clustering coefficient $C$, long time behaviour.  The left panel shows a color plot covering the parameter space, in which a dashed line (defined as the contour corresponding to $R=1.5$) separates the single species region  (``single sp.'') from the speciation regions. The right panel presents $C$ vs. $B$ for some specific values of the parameter  $S$.   }
    \label{fig:clus_asymp}
\end{figure}

\begin{figure}[h!]
   \centering
    \makebox[\textwidth][c]{%
    \begin{subfigure}[b]{0.528\textwidth}
   \includegraphics[width=\linewidth]{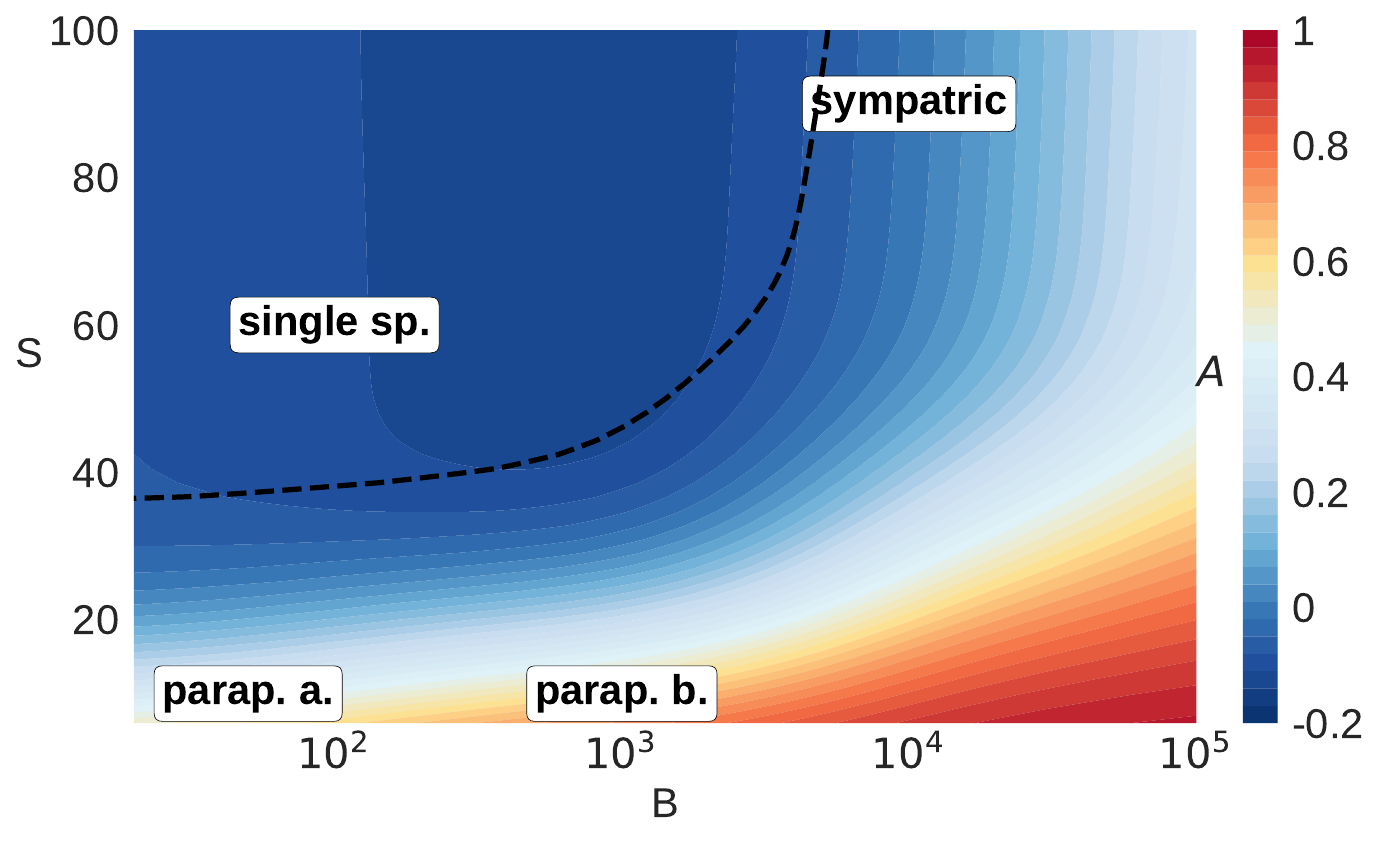}
    \end{subfigure}%
    \hspace{0.5em}\begin{subfigure}[b]{0.624\textwidth}
        \includegraphics[width=\linewidth]{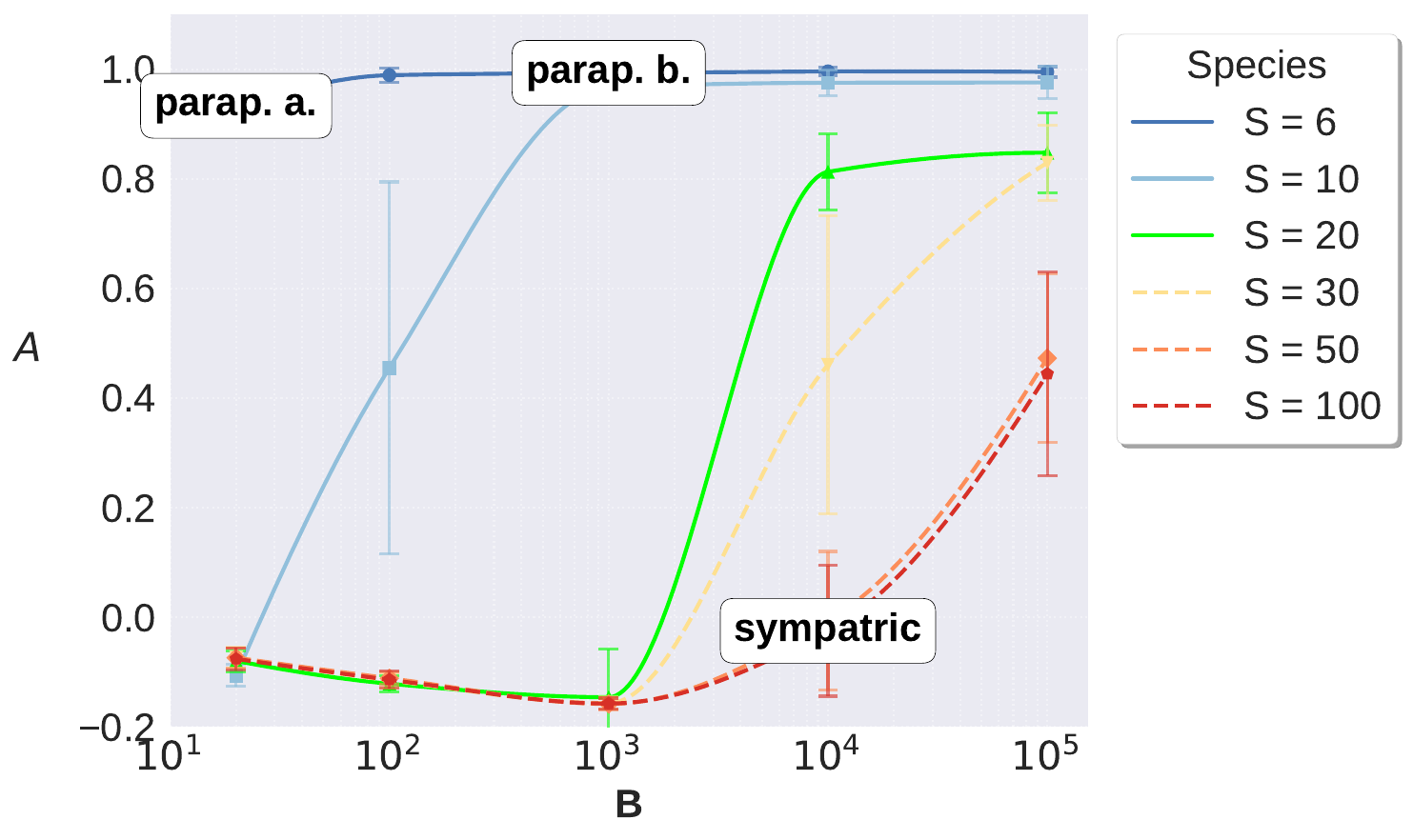}
   \end{subfigure}}
  \caption{Assortativity coefficient $A$, long time behaviour.   The left panel shows a color plot covering the parameter space, in which a dashed line  (defined as the contour corresponding to $R=1.5$) separates the single species region  (``single sp.'') from the speciation regions. The right panel presents $A$ vs. $B$ for some specific values of the parameter  $S$.   }
    \label{fig:assort_asymp}
\end{figure}

\begin{figure}[h!]
   \centering
    \makebox[\textwidth][c]{%
    \begin{subfigure}[b]{0.528\textwidth}
   \includegraphics[width=\linewidth]{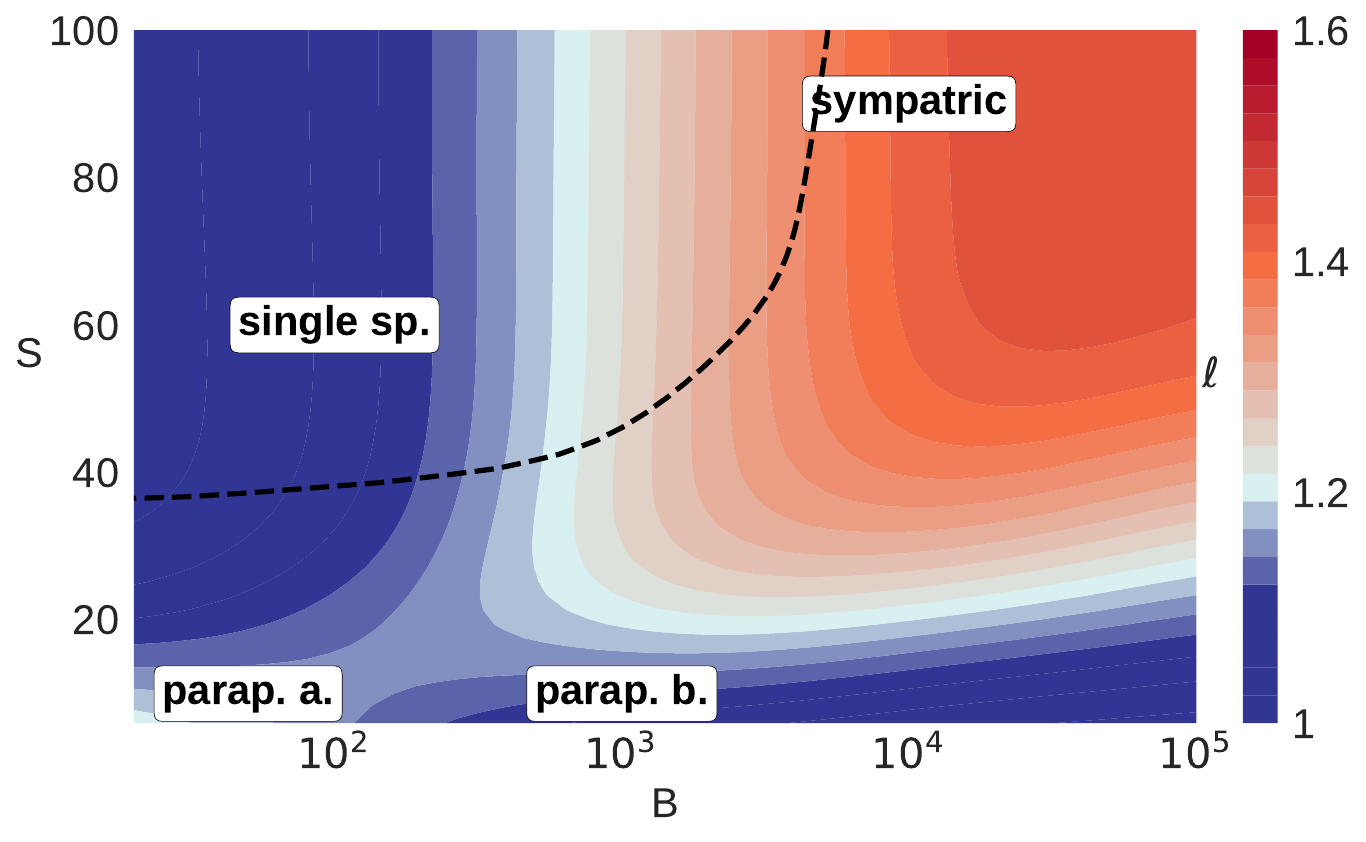}
    \end{subfigure}%
    \hspace{0.5em}\begin{subfigure}[b]{0.624\textwidth}
        \includegraphics[width=\linewidth]{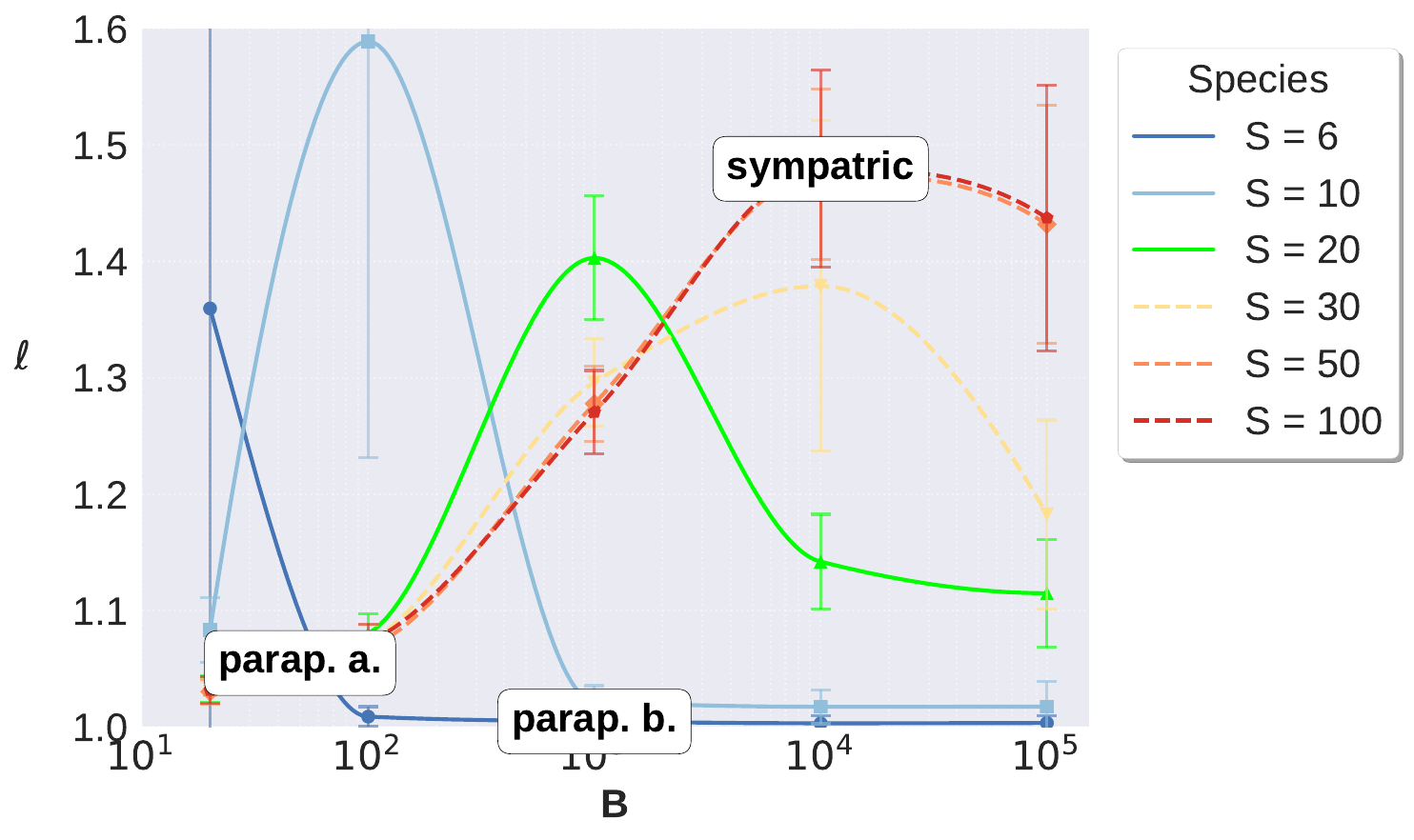}
   \end{subfigure}}
  \caption{Mean geodesic path $l$, long time behaviour. The left panel shows a color plot covering the parameter space, in which a  dashed line (defined as the contour corresponding to $R=1.5$) separates the single species region  (``single sp.'') from the speciation regions. The right panel presents $l$ vs. $B$ for some specific values of the parameter  $S$.  }
    \label{fig:shPh_asymp}
\end{figure}

Figures \ref{fig:rich_asymp} to \ref{fig:shPh_asymp} display the density distribution across the entire parameter space for each structural network property (left panels). The right panels show curves illustrating the dependence of these properties on $B$ for specific values of $S$.

The main observations derived from these figures are:

\begin{enumerate}
    \item Speciation can occur even under full sympatry ($S = 100$ in our simulations, which emulate the maximum distance between points on the torus) when genomes reach sufficiently large sizes. This finding is consistent with the Higgs-Derrida theory \cite{Higgs-Derrida}, extensively analyzed in Ref. \citenum{Aguiar2017}.

    \item Parapatric speciation patterns (a) and (b) are clearly distinguishable: pattern (a) comprises a localized region of very high richness ($R \simeq 100$, for $S \lesssim 10$ and $B \lesssim 100$), while pattern (b) covers a more extensive region of intermediate richness (small $S$ and $B> 100$).

    \item A fourth region emerges for high $S$ and small $B$ where speciation does not occur (single species region, $R = 1$).

    \item The giant component size (Fig. \ref{fig:giant_asymp}) is directly linked to richness: it encompasses the entire network in the single-species region, but decreases significantly during speciation events following the fission phase.

    \item Network connectivity (Fig. \ref{fig:conn_asymp}) follows a similar pattern, suggesting that clusters develop more internal links when the network is dominated by the giant component.

    \item While GSN under parapatric conditions exhibit high local clustering, networks from sympatric scenarios are nearly indistinguishable from single-species region networks in terms of this metric (Fig. \ref{fig:clus_asymp}).

    \item Networks show high degree assortativity in speciation regions (Fig. \ref{fig:assort_asymp}). This effect is enhanced by physical mating restrictions, particularly for small $S$ values where strong assortativity persists for most genome sizes. Conversely, the single-species region displays negative assortativity, and sympatric speciation approaches the behavior of random networks \cite{entropy_2025} in terms of assortativity.

    \item The mean geodesic path (Fig. \ref{fig:shPh_asymp}) shows striking correlation with local clustering and helps distinguish between sympatric and parapatric speciation. With the exception of the marginal parapatric scenario (a) discussed previously ($S \lesssim 10$ and $B \lesssim 100$), where mean geodesic paths reach up to 4 units, the post-transient mean geodesic path remains below 1.5 for all other cases.
\end{enumerate}

It is important to emphasize that our networks are constructed based on Hamming distances between binary sequences. Therefore, these values are not directly related to the spatial parameters $S$ and $L$. A complementary study constructing networks based on real space considerations is currently in progress.

\section{Final Remarks and Conclusions}\label{sec:conclusions}

The present article demonstrates that genetic similarity networks (GSN) are an effective tool to distinguish between sympatric and parapatric speciation modes. By analyzing structural properties of these networks -such as the local clustering coefficient, average degree, assortativity, and geodesic path length— clear and distinctive patterns are identified for each  speciation type. In the following paragraphs we outline the most outstanding findings of our work.

\subsection{Temporal behavior}
 On the first hand, it is important to stress that  when genomes reach sufficiently large sizes, we obtain speciation in full sympatric conditions. In this case, the onset of the  speciation events  occurs at a similar number of generations than for parapatric scenarios, even when the characteristic of the temporal course of richness and giant component size is different in comparison with this case.
Additionaly, we identify four distinct regions in the parameter space: one localized region with extremely  high richness (very small $S$ and $B$), another extensive area with intermediate richness (small $S$ and intermediate to high $B$), a third  region with low richness (high $S$ and high $B$), and a region lacking of richness ($R=1$, high $S$ and small to intermediate $B$). These correspond to parapatric (a.), parapatric (b.), sympatric, and single-species scenarios, respectively. All these findings support the validity of our modification  of the original de Aguiar model.

 In what concerns temporal evolution, after a single-species phase of around $10^2$ generations, speciation is followed by two differentiated  phases. A fast initial phase, in which  network metrics evolve to a stationary state, followed by a slow phase in which only the richness and the giant component exhibit changes. In sympatry, fast speciation is abrupt and has a very low richness, contrary to the parapatric case, in which the fission phase  extends  over many generations and gives rise to the formation of many species. The slow phase has a constant slope in parapatric scenarios, which is different for each set of parameters but it  seems to accelerate after $10^3$ generations in the sympatric configuration.

 \subsection{Post-transient states}

 Networks properties at quasi-stationary regime  are calculated at generation $2000$ in the whole range of $S$ and for genome sizes up to $10^5$. In section \ref{subsec:evolution_struct_props} we check that all the metrics stabilize at  this time, except for the richness and the giant component, which define the different speciation phases.
 The main  observations are: (1.) Parapatric speciation is characterized by networks exhibiting high assortativity and high local clustering, with mean geodesic paths extremely  sensitive to the genome size, with values close to 1 in almost the whole range. We remark  the  exception observed  for very small values of the parameter $B$, where mean  geodesic path length  can increase beyond 4 units. In sympatric speciation, on the other hand, assortativity takes values close to zero, indicating that these networks are much more disordered, and  the mean  geodesic path stabilizes around the value $1.5$.  These last outcomes resemble  artificially generated random networks \cite{entropy_2025}.
 In addition, local clustering is consistently below one ($\sim 0.75$ for large genomes). Remarkably, the single-species region shows negative assortativity (disassortative networks), while its clustering coefficients and mean geodesic path are indistinguishable from parapatric values. (2.) Structurally, parapatric networks exhibit very low connectivity at post-transient regime. This fact contrasts with the behavior of geodesic paths (Figure \ref{fig:conn_asymp}), which remain close to 1, indicating an apparent high connectivity. The reason for this discrepancy is that geodesic path length represents an average over all clusters which, in parapatric cases, have a high density of links. Nevertheless, the loss of connections between the different clusters due to fission causes connectivity to drop sharply,  overriding the effect of the dense within-clusters connections. By contrast, sympatric clusters exhibit sparser internal connections than their parapatric counterparts, and yet the global connectivity remains close to 1/2. This is because, in this case, global connectivity essentially reflects the connectivity of the giant component, whose structure is approximately that of a random network.  (3.)
 Parapatric escanarios (a.) and (b.) identified in Refs. \cite{costa_1, costa_2}, are also resolved through network analysis. Moreover,  our simulations  indicate that parapatric scenario (a.) is restricted to a narrow portion of the $S-B$ plane, in contrast to parapatric region (b.), which occupy almost all the parapatric  region. (4.) In all the simulated cases, sympatric conditions exhibit much higher fluctuations and require therefore, a significantly larger number of realizations. This statistical fact  —also reported by Costa et al. \cite{costa_1,costa_2}— arises  as a direct consequence of the sample, size relative to the size of the space of results.

From the previous discussion  it can be concluded that, while local clustering and assortativity appear to be promising candidates for identifying speciation modes,  geodesic paths have a modest value, as this metric does not  distinguish between single-species and parapatric regions.
In the Supplementary Material we present a preliminary Principal Analysis supporting these assertions.

\subsection{Biological interpretation}

    In parapatric speciation, high clustering coefficients combined with low mean degrees point to strong reproductive cohesion within spatially restricted groups. In the sympatric regime, by contrast, wider spatial distribution promotes greater dispersion in the  genotypic space, which in turn reduces intraspecific connections more markedly. The mean degree, however, is less affected overall, as it is driven primarily by the loss of connections resulting from speciation itself. These patterns are further supported by geodesic path length behavior: short mean paths in parapatric networks suggest tight inbreeding and fragmented population structure, whereas the slightly longer paths observed under sympatry reflect a more even distribution of genetic diversity, with no pronounced subdivisions.

    Assortativity is especially high in parapatric scenarios, signalling active assortative mating among genetically similar individuals, a process that directly fosters reproductive isolation and divergence.

    Lastly, the observed ``fission'' versus ``evaporation'' patterns in the richness and the giant component dynamics correspond to two distinct micro-evolutionary regimes. Fission mirrors allopatric-like fragmentation into daughter groups, while evaporation resembles peripatric speciation through the continuous loss of small clusters from a large, stable central population. This interpretation is consistent with Higgs–Derrida theories of neutral speciation\cite{Higgs-Derrida,Aguiar2017,costa_2}

\subsection{Further investigations}

The investigation presented in this work opens many avenues that would be highly interesting to explore in future studies. These include:

\begin{enumerate}
    \item[(a)] A thorough quantitative comparison between our adaptation of the de Aguiar model and the original version. In this respect, it would be interesting to incorporate the movement of individuals in our implementation of the model.

    \item[(b)] The long-term behavior of richness and the giant component under sympatric conditions.

    \item[(c)] The damped oscillations in connectivity and clustering observed at the onset of speciation. A satisfactory explanation of this phenomenon would require new and highly specific simulations.
\end{enumerate}

 Furthermore, the results presented along this work give rise to the following interesting   questions.
For instance, are the observed properties of the GSN compatible with established models of network generation? Specifically, are GSN best described as random networks (Erdős–Rényi \cite{random-graphs}),
 scale-free networks (Barabási–Albert \cite{free-scale}),
 small-world networks (Watts–Strogatz \cite{small-world}),
 geographical network, regular networks, or a hybrid of these?
 And finally, how do our results change in the presence of geographical barriers (allopatric speciation)? Is there a critical barrier size that induces a qualitative shift in the behavior of the observables?

\section*{Acknowledgments}

 C. P.  thanks CONICET for partial support of this work through grant PIP11220210100546CO. CNEA and SECYT are also gratefully acknowledged for providing  supercomputing resources for our LabSim Supercomputing Lab.
 D. M. S. thanks CONICET for its support through grant PIP 112202201001260CO.

\bibliographystyle{apsrev4-2}
\bibliography{gsn.bib}

\section*{Appendix A}

In this appendix we introduce the  mathematical  definitions
of the measures and observables studied in this article.
Unless otherwise specified, the summation indices run from 1 to the network size $N$.

 \subsection*{Hamming distance}
Let $\sigma_1$ and $\sigma_2$ denote two binary strings of length $B$, namely:
\[
\sigma_1=(\sigma_1(1),\sigma_1(2),\ldots,\sigma_1(B))
\]
 and
 \[
\sigma_2=(\sigma_2(1),\sigma_2(2),\ldots,\sigma_2(B)),
\]
 where $\sigma_i(k)\in\{0,1\}$ for $i=1,2$ and $k=1,2,\ldots,B$. The Hamming distance between the two strings is defined as:
 \[
 H(\sigma_1,\sigma_2)=\sum_{k=1}^B(\sigma_1(k)-\sigma_2(k))^2.
 \]
 According to this definition, pairs of strings of length $B$ can have any integer distance between $0$ and $B$.

 \subsection*{Connectivity}

 This quantity also corresponds (for large networks)  to the averaged normalized degree, and is defined as:
 \[
 D=\frac{\sum_{ij}a_{ij}}{N(N-1)}=\frac{1}{N(N-1)}\sum_i k_i\simeq \frac{\left<k\right>}{N}.
 \]
 where $a_{ij}$ are the elements of the adjacency matrix of the graph associated to the network,
 \[
 a_{ij}=\begin{cases}
1 & \text{if there is a link between nodes }i\;  {\rm and \;} j \\
0 & \text{otherwise,}
\end{cases}
\]
and $k_i$ the degree of the node $i$:
\[
k_i=\sum_j a_{ij}.
\]

\subsection*{Local clustering}
Local clustering coefficient measures the  fraction of neighbors of a given node that are also connected to each other; or equivalently,  the fraction of possible links between the neighbors of a node that are actually present in the network. The formal definition is:

\[
C_i = \frac{T_i}{k_i (k_i - 1) / 2},
\]

where
\[
T_i = \frac{1}{2} \sum_{jk} a_{ij} a_{jk} a_{ki}
\]
is the number of triangles that exist in the network to which node $i$ belongs. From this we obtain $C$ by simply averaging  $C_i$ over the $N$ nodes.
\[
C=\frac{1}{N}\sum_i C_i.
\]

\subsection*{Degree assortativity}
In our simulations we employ Holme  formula for assortativity \cite{holme_2007}:

\begin{equation}
A = \frac{4\langle k_1 k_2\rangle - \langle k_1 + k_2\rangle^2}{2\langle k_1^2 + k_2^2\rangle - \langle k_1 + k_2\rangle^2}.
\label{eq:r}
\end{equation}
In this formula, $k_1$ and  $k_2$ correspond  to the degrees of the two  nodes connected by a given link, and $\langle\ldots\rangle$ represents an average over all links present in the network. In undirected networks, both $k_1$ and $k_2$ follow  the same probability distribution, and we have therefore the following identities:
\[
\langle k_1 + k_2\rangle= 2\langle k_1\rangle\equiv 2\bar{k},
\]
\[
\langle k_1^2 + k_2^2\rangle=2\langle k_1^2 \rangle\equiv  2\overline{k^2}.
\]
Accordingly, equation (\ref{eq:r}) reduce to

\begin{equation}
A = \frac{4\langle k_1 k_2\rangle -4(\bar{k})^2}{4\overline{k^2}-4(\bar{k})^2}
=\frac{\langle k_1 k_2\rangle -\langle k_1\rangle \langle k_2\rangle}{\sqrt{\langle k_1^2\rangle-\langle k_1\rangle^2}\sqrt{\langle k_2^2\rangle-\langle k_2\rangle^2}}
=\frac{{\rm Cov}(k_1,k_2)}{\sqrt{{\rm Var}(k_1)}\sqrt{{\rm Var}(k_2)}},
\end{equation}
which is precisely the Pearson correlation coefficient of the degrees of the two nodes connected by a link.
\subsection*{Geodesic path}

The geodesic path length $G_{ij}$ represents the minimum number of links that must be traversed to reach node $j$ starting from node $i$. We deal here with undirected networks, so this coefficient is symmetric. In terms of the adjacency matrix, it corresponds to the minimum integer $n$ such that
\[
(a^n)_{ij} > 0.
\]

We obtain $G$ by averaging over all pairs of nodes for which $G_{ij}<\infty$ (namely, pairs of nodes belonging to the same connected component).

\clearpage
\section*{Supplementary Material}
\setcounter{figure}{0}
\setcounter{table}{0}
\renewcommand{\thefigure}{S\arabic{figure}}
\renewcommand{\thetable}{S\arabic{table}}

\subsection*{Long-time simulations of the Sympatric case}

\begin{figure}
\centering
\includegraphics[width=0.7\columnwidth]{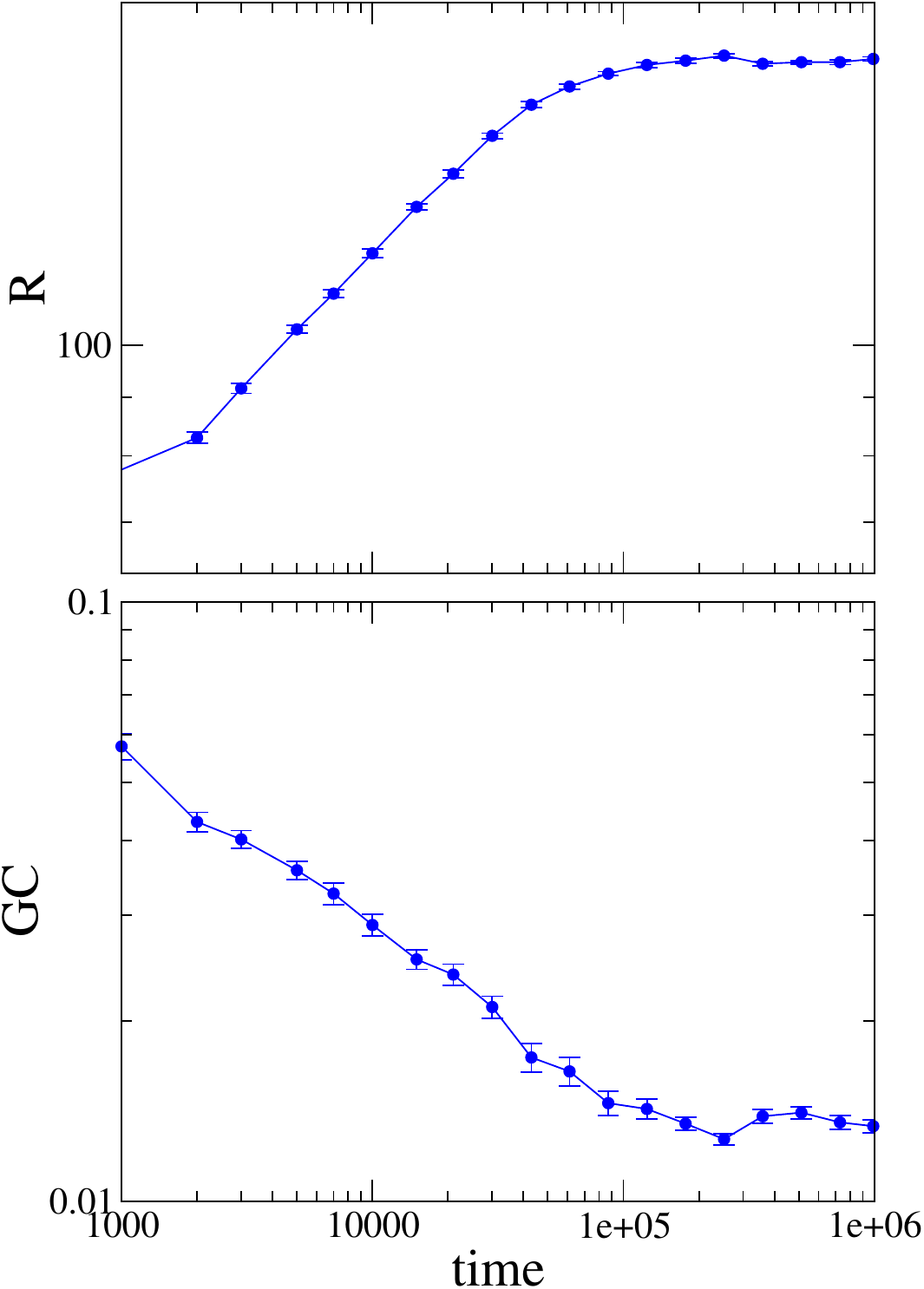}
\caption{ Richness and giant component for a long-time simulation corresponding to the Parapatric a. case. The asymptotic behavior of both quantities is achieved, by spanning
	6 orders of magnitude of the evolutionary process.
    \label{R_GC_long}
    }

\end{figure}

We performed a very long simulation spanning six orders of magnitude  to analyze the asymptotic parapatric a.  regime.
In Figure \ref{R_GC_long} we show  the richness  \(R\) (upper panel) and the giant component \(GC\) (lower panel). The speciation process reaches a clear plateau   not observed in Figure 6 of the manuscript, which presents a simulation  time of 50000 steps.

\subsection*{Robustness against genetic threshold, population size, and mutation rate}

\begin{figure}[htbp]
    \centering
    \begin{subfigure}{0.45\textwidth}
        \centering
        \includegraphics[width=\textwidth]{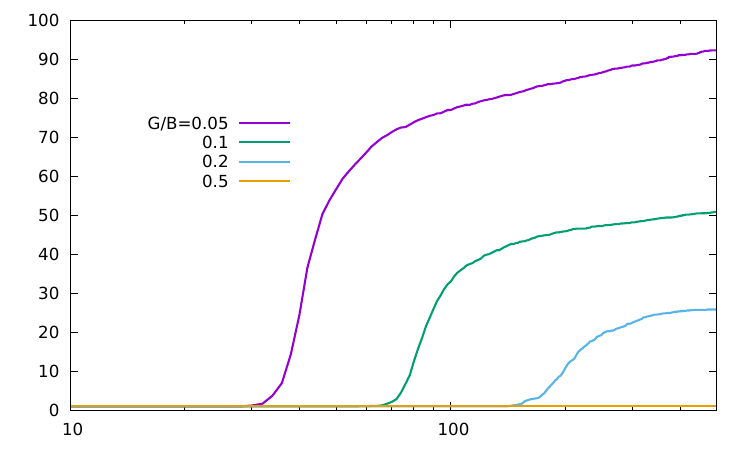}
        \caption{Richness, for several ratios $G/B$.}
        \label{fig:panel1}
    \end{subfigure}
    \hfill
    \begin{subfigure}{0.45\textwidth}
        \centering
        \includegraphics[width=\textwidth]{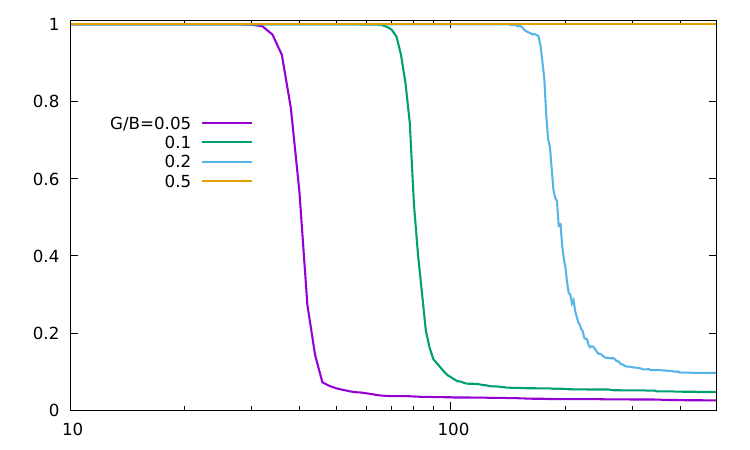}
        \caption{GC, for several ratios $G/B$.}
        \label{fig:panel2}
    \end{subfigure}

    \vspace{0.5cm}

    \begin{subfigure}{0.45\textwidth}
        \centering
        \includegraphics[width=\textwidth]{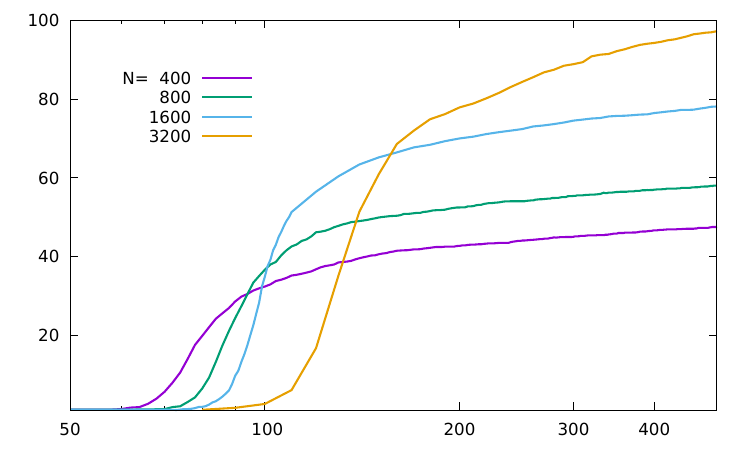}
        \caption{Richness, for different population sizes}
        \label{fig:panel3}
    \end{subfigure}
    \hfill
    \begin{subfigure}{0.45\textwidth}
        \centering
        \includegraphics[width=\textwidth]{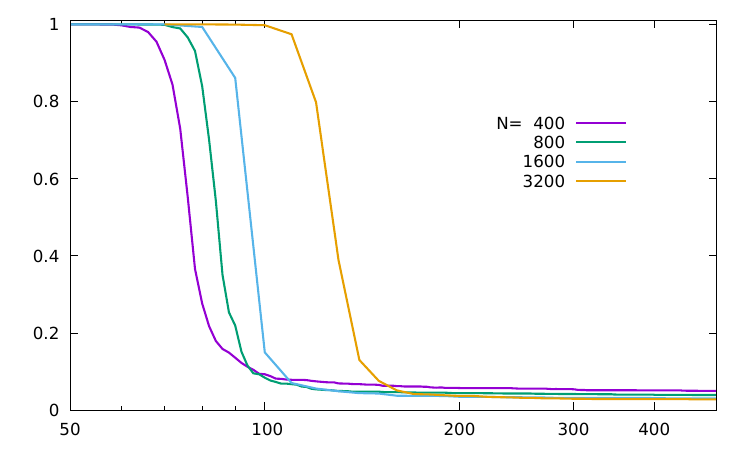}
        \caption{GC, for different population size.}
        \label{fig:panel4}
    \end{subfigure}

    \vspace{0.5cm}

    \begin{subfigure}{0.45\textwidth}
        \centering
        \includegraphics[width=\textwidth]{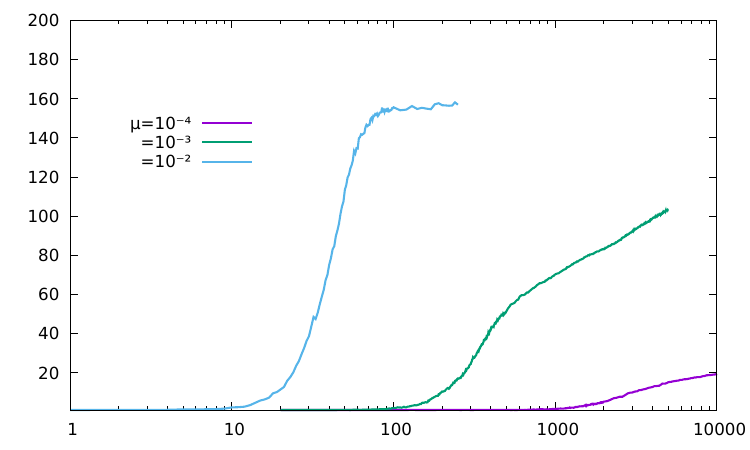}
        \caption{Richness, for different mutation rates.}
        \label{fig:panel5}
    \end{subfigure}
    \hfill
    \begin{subfigure}{0.45\textwidth}
        \centering
        \includegraphics[width=\textwidth]{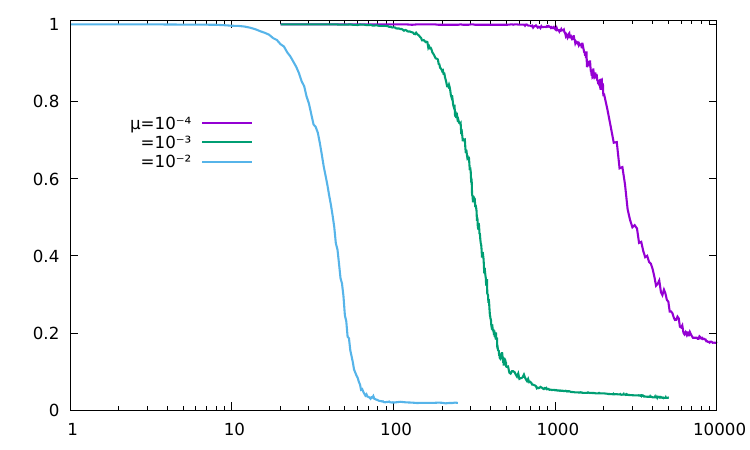}
        \caption{GC, for different mutation rates.}
        \label{fig:panel6}
    \end{subfigure}

    \caption{Behavior of the richness and the giant component as the genetic threshold, the population size, and the mutation rate go beyond the typical values employed in the main paper.}
    \label{robustness_G-N-mu}
\end{figure}

In Figure \ref{robustness_G-N-mu}
we present the simulation results for  richness and giant component as a function of time for a set of values of $G/B$, a set of values of $N$ and a set of values of $\mu$.

\subsection*{Speciation-diagram}
\begin{figure}[ht!]
    \centering
    \makebox[\textwidth][c]{%
    \begin{subfigure}[b]{0.6\textwidth}
   \includegraphics[width=\linewidth]{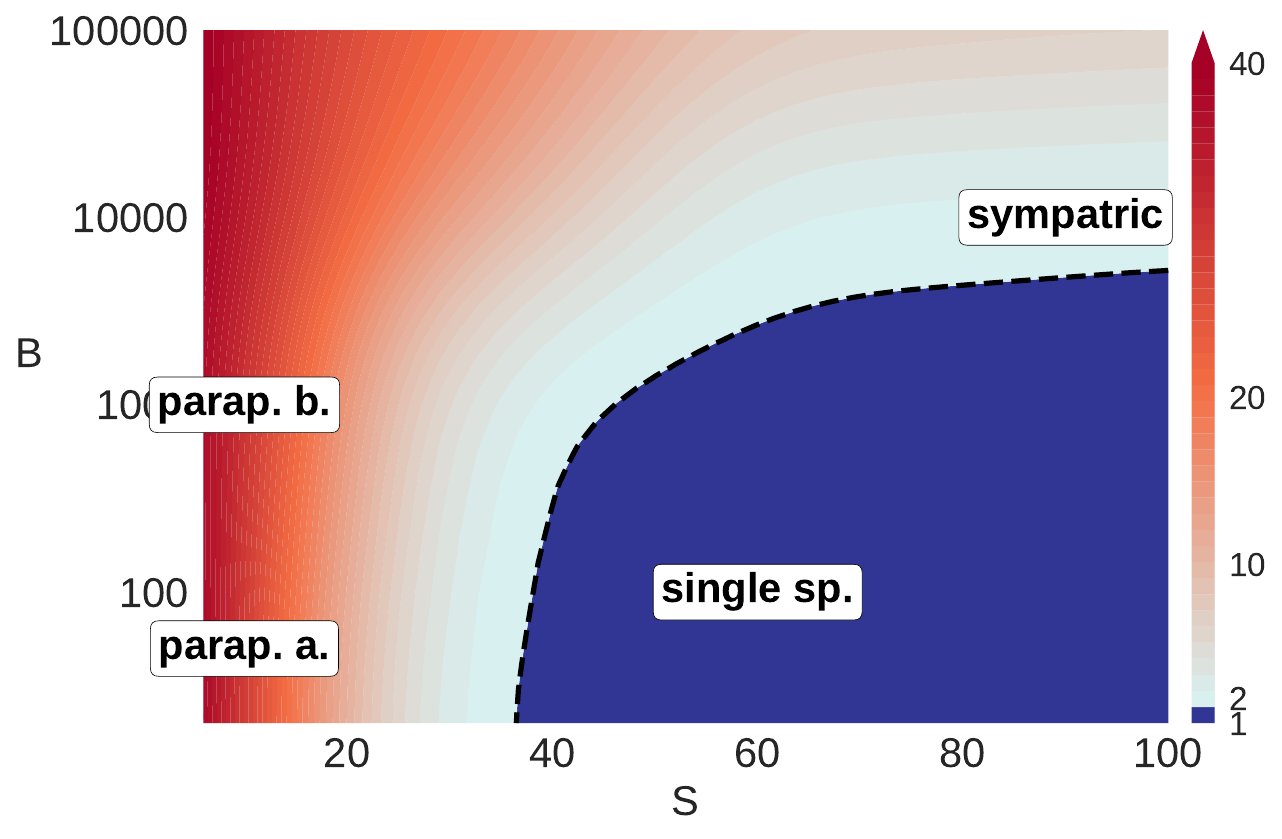}
    \end{subfigure}%
    \hspace{0.5em}\begin{subfigure}[b]{0.52\textwidth}
        \includegraphics[width=\linewidth]{./speciation_diagram_aguiar.pdf}
   \end{subfigure}}
\caption{Left Panel: Full speciation  diagram  as obtained from our simulations. Right Panel: speciation diagram taken from de Aguiar\cite{Aguiar2017}.\label{fig:full_phDiag}}
\end{figure}

Fig. \ref{fig:full_phDiag} presents a  full phase diagram depicting every region of speciation in the post-transient regime (see section III. B). For comparison, we place alongside this plot, the corresponding speciation diagram obtained by de Aguiar in Ref. \citenum{Aguiar2017}.

\subsection*{Principal Component Analysis (PCA)}
\begin{figure}[ht!]
    \centering
    \includegraphics[width=\linewidth]{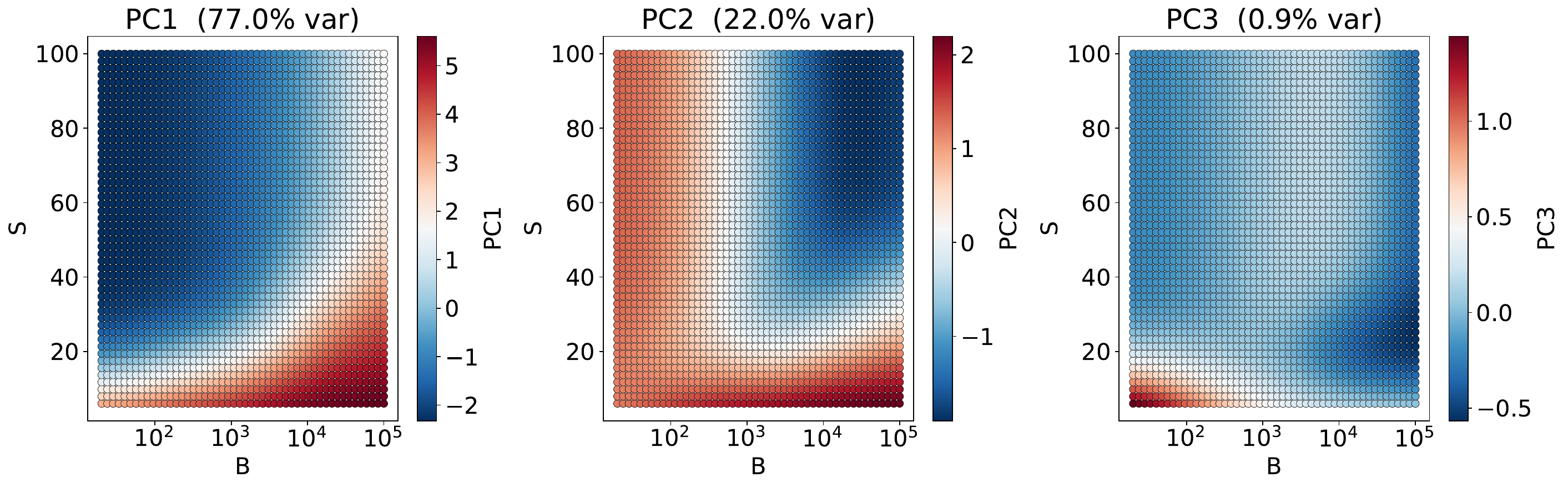}
\caption{\label{fig:PCA} The three principal components and their respective variance contribution.}
\end{figure}

    Fig. \ref{fig:PCA}  shows the heat maps corresponding to the first three principal components. The left panel corresponds to  PC1 (mainly represented by $R$, $GC$, $D$, $C$, and $A$; see Table 1), which separates the parapatric regions from the single-species and sympatric regions. The middle panel shows PC2 (whose main contribution comes from the geodesic path, followed by $R$, with almost no contributions from $GC$, $D$, $C$, and $A$), separating the area of sympatric speciation from single-species and parapatric speciation. The PC3 component, representing just about 1\% of the variance and exhibited in the right panel, resolves the parapatric a. region. Accordingly, the initial separation of regions adopted by Costa {\it et al} \cite{costa_1,costa_2} acquires, through this analysis, a firm quantitative basis.

\begin{table}[htbp]
\centering
\caption{Variance contribution of the principal components.}
\label{tab:variance}
\begin{tabular}{lcc}
\hline
\textbf{Component} & \textbf{Variance contribution (\%)} & \textbf{Cumulative (\%)} \\
\hline
PC1 & 76.97 & 76.97 \\
PC2 & 21.95 & 98.93 \\
PC3 &  0.94 & 99.87 \\
PC4 &  0.12 & 99.99 \\
PC5 &  0.01 & 100.00 \\
PC6 &  0.00 & 100.00 \\
\hline
\end{tabular}
\end{table}

\begin{table}[htbp]
\centering
\caption{Loadings matrix of the principal components.}
\label{tab:loadings}
\begin{tabular}{lcccccc}
\hline
 & \textbf{PC1} & \textbf{PC2} & \textbf{PC3} & \textbf{PC4} & \textbf{PC5} & \textbf{PC6} \\
\hline
$R$    &  0.4184 &  0.3394 &  0.8418 &  0.0281 & -0.0196 &  0.0000 \\
$GC$   & -0.4555 & -0.1676 &  0.2772 & -0.0762 & -0.8257 &  0.0000 \\
$D$    & -0.4478 &  0.2334 &  0.1185 &  0.4374 &  0.1990 &  0.7071 \\
$C$    & -0.4478 &  0.2334 &  0.1185 &  0.4374 &  0.1990 & -0.7071 \\
$A$    &  0.4554 &  0.1574 & -0.3230 &  0.6762 & -0.4541 &  0.0000 \\
$\ell$ &  0.0952 & -0.8503 &  0.2866 &  0.3916 &  0.1801 &  0.0000 \\
\hline
\end{tabular}
\end{table}

\begin{figure}[ht!]
    \centering
    \includegraphics[scale=.7]{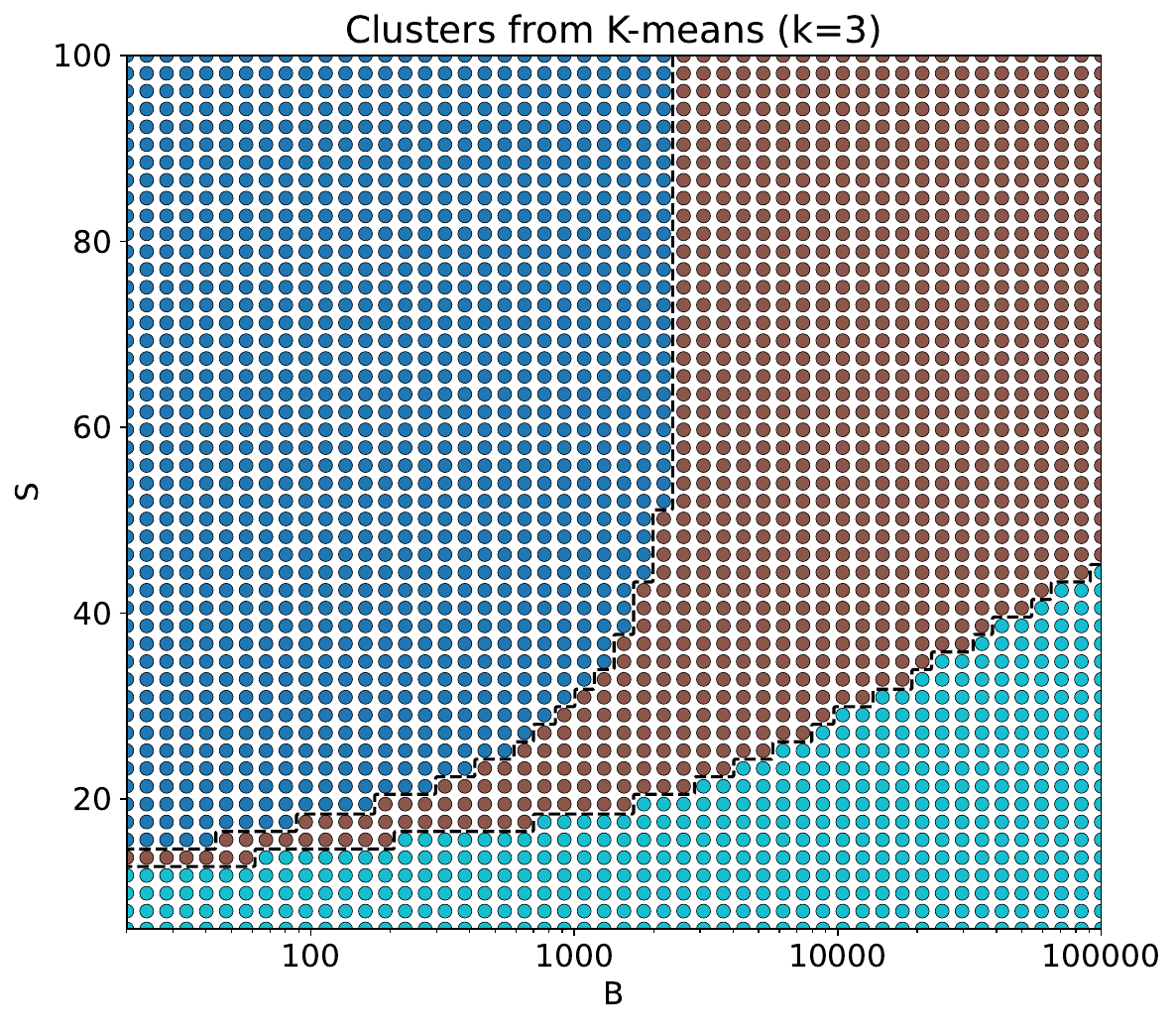}
\caption{\label{fig:regions} Principal Component Analysis (PCA): different clusters obtained through the method of K-means.}
\end{figure}

Figures \ref{fig:regions} displays the final regions obtained from the PCA, after implementing the K-means approach with the  ``elbow'' criterion.  We notice that this method does not resolve parapatric sub-regions.

\end{document}